\documentclass[aps,prd,a4paper,onecolumn,amsmath,showpacs,superscriptaddress,nofootinbib,preprintnumbers,notitlepage]{revtex4-1}
 
\usepackage{verbatim}
\usepackage[T1]{fontenc}
\usepackage[utf8]{inputenc}
\usepackage[american]{babel}
\usepackage{graphicx,subcaption,caption}
\usepackage{epsfig}
\usepackage{graphicx,subcaption,caption}
\usepackage{pgfplots}
\usepackage{booktabs}
\usepackage{multirow}
\usepackage{dcolumn}
\usepackage{amsmath}
\usepackage{mathtools}
\usepackage{amsfonts}
\usepackage{amssymb}
\usepackage{epstopdf}
\usepackage{bm}
\usepackage{siunitx}
\usepackage{braket}
\usepackage{enumitem}
\usepackage{soul}
\usepackage{color}
\usepackage{amsmath}

\usepackage{transparent}
\usepackage{pifont}

\definecolor{navyblue}{rgb}{0.0, 0.0, 0.5}
\definecolor{royalblue}{rgb}{0.25, 0.41, 0.88}
\definecolor{cadmiumgreen}{rgb}{0.0, 0.42, 0.24}
\definecolor{blue-violet}{rgb}{0.54, 0.17, 0.89}
\definecolor{darkviolet}{rgb}{0.58, 0.0, 0.83}
\definecolor{orange(colorwheel)}{rgb}{1.0, 0.5, 0.0}

\usepackage{hyperref}
\hypersetup{
    colorlinks=true, 
    linkcolor=royalblue, 
    citecolor=magenta}

\newcommand\be{\begin{equation}}
\newcommand\ee{\end{equation}}

\newcommand\bea{\begin{eqnarray}}
\newcommand\eea{\end{eqnarray}}

\usepackage{booktabs}
\usepackage{multirow}
\usepackage{dcolumn}
\usepackage{colortbl}

\definecolor{magenta(process)}{rgb}{1.0, 0.0, 0.56}

\definecolor{darkspringgreen}{rgb}{0.09, 0.45, 0.27}

\definecolor{royalblue(web)}{rgb}{0.25, 0.41, 0.88}

\begin{document}

\title{Slow-roll inflation in $f(Q)$ non-metric gravity}

\author{Salvatore Capozziello}
\email{capozziello@na.infn.it}
\affiliation{Dipartimento di Fisica "E. Pancini", Universit\'a di Napoli Federico II", Via Cinthia, I-80126, Napoli, Italy}
\affiliation{Istituto Nazionale di Fisica Nucleare (INFN), sez. di Napoli, Via Cinthia 9, I-80126 Napoli, Italy}
\affiliation{Scuola Superiore Meridionale, Largo S. Marcellino, I-80138, Napoli, Italy}

\author{Mehdi Shokri}
\email{mehdishokriphysics@gmail.com}
\affiliation{School of Physics, Damghan University, P. O. Box 3671641167, Damghan, Iran}
\affiliation{Department of Physics, University of Tehran, North Karegar Ave., Tehran 14395-547, Iran}
\affiliation{Canadian Quantum Research Center 204-3002 32 Avenue Vernon, British Columbia V1T 2L7 Canada}

\preprint{}
\begin{abstract}
We discuss the cosmological  inflation in the context of $f(Q)$ non-metric gravity, where $Q$ is the non-metric scalar. After introducing conformal transformations for $f(Q)$ gravity, we first focus on the potential-slow-roll inflation by studying the corresponding potentials for different forms of the function $f(Q)$ in the Einstein frame. Secondly, we investigate the Hubble-slow-roll inflation for three classes of inflationary potentials considered for the specific form $f(Q)\propto Q^{2}$, in the Jordan frame.  We compare results in both approaches with observations coming from Planck and BICEP2/Keck array satellites.  Observational constraints on the parameters space of the models are obtained as well. 
\end{abstract}
\date{\today}
\maketitle
\section{Introduction}\label{intro}
Observational evidences of late-time  acceleration  opened  new windows in modern cosmology due to many attempts proposed to explain it. The phenomenon is generically dubbed Dark Energy (DE) but the possibility to reconstruct a coherent and self-consistent cosmic history, working at any epoch, is far to be easily achieved.  Cosmological constant $\Lambda$ is the straightforward explanation which should work at early (inflation) and late (dark energy) epoch.  The approach navigates us to the $\Lambda$CDM model that claims the universe is formed by $\sim72\%$ DE, $\sim24\%$ Dark Matter (DM), and $\sim4\%$ visible matter, requiring an  early inflationary epoch capable of solving the shortcomings of Cosmological Standard Model based on General Relativity, Big Bang Nucleosynthesis  and Standard Model of Particle. Despite the successes of this coarse grained model, it suffers from a vital ambiguity that is related to the existence of a big difference between the value of $\Lambda$ predicted by any quantum gravity theory  and the present  observational value. Furthermore, there are no final experimental evidence of new  particles capable of explaining DM and DE at fundamental level. In this situation, considering the gravitational counterpart can be a reasonable way out to cure shortcomings and address phenomenology.  In this perspective, extensions and modifications of GR seem viable approaches both at early and late epochs. See for example \cite{Clifton:2011jh,Capozziello:2011et,
Faraoni:2010pgm, Cai:2015emx,Nojiri:2017ncd, Saridakis:2021vue,CANTATA:2021ktz}. 

Besides cosmology, these research activities lead also to some fundamental questions related to the  representation, the number of degrees of freedom and the true physical interpretation  of the gravitational field.

For example, in the framework of GR,  the Ricci curvature scalar $R$ gives rise to the spacetime dynamics. It is derived from   the Levi-Civita connection   fixed by the Equivalence Principle. In this context, gravity is a metric theory where the Lorentz invariance,  causality and other well-established principles hold. 

However, we can represent gravitational field by other approaches based on further geometric quantities   like torsion and non-metricity. 
These  representations are dynamically equivalent to GR but some basic principles are completely different. 
For example, the so-called teleparallel equivalent of general relativity (TEGR) assumes  torsion as the field describing gravity.   Here  curvature and non-metricity are zero and one adopts the Weitzenb\"{o}ck connection as the affine connection  \cite{1,Bahamonde:2021gfp}. In this case, the fundamental objects are tetrads by which it is possible to derive the affine connection, the torsion invariants and finally the field equations.

Another equivalent formulation of GR is based on non-metricity. Here,  a flat spacetime  can  be considered and gravitational field  is associated to non-metricity. This formulation is known as the symmetric teleparallel equivalent of GR  (STEGR) \cite{Nester:1998mp,Adak:2005cd,Adak:2008gd,Adak:2004uh,3,Jarv:2018bgs}. In STEGR there is a metric tensor,  geometry is equipped  with a non-metric connection 
but total curvature and  torsion are vanishing.

Like curvature can be related to  rotation of vectors when parallel transported on  closed curves, non-metricity can be related to the change of vector lengths  when parallel transported and then gravity can be dealt under the standard of  gauge theories.

The mentioned equivalent approaches are often named "The Geometrical Trinity of Gravity" \cite{BeltranJimenez:2019esp,Capozziello:2021pcg}. 

Although these three versions of GR are fully equivalent from a dynamical point of view, the basic principles on which they are formulated are very different. Furthermore, their modifications are not  equivalent at all being  the theories  formulated in different geometries. Hence, one can find a wide range of possibilities and interpretations that have to be consistently compared with observational and experimental data. For example, also if GR and TEGR are dinamically equivalent, this is not the case for   $f(R)$ gravity \cite{4,Capozziello:2011et} and $f(T)$ gravity \cite{Cai:2015emx,6,7,8}. The first is a fourth-order theory in metric formulation, the latter remain a second-order theory like GR and TEGR. 
 
Another possible extension is 
 $f(Q)$ gravity \cite {BeltranJimenez:2017tkd} constructed by extending STEGR which is the particular case $f(Q)=Q$. 
 Such a theory, not requiring, {\it a priori},  the Equivalence Principle, is suitable to be dealt, as said,  under the standard of gauge theories and presents other advantages. For example, it 
  seemingly shows not strong coupling problems because of additional scalar modes \cite{9} while the $f(T)$ gravity presents these problems when perturbations around a Friedman-Lema\^itre-Robertson-Walker (FLRW) metric are considered \cite{10}. Also, the linear perturbations of scalar, vector, and tensor modes in $f(Q)$ gravity have been studied  \cite{BeltranJimenez:2019tme}. 
  
  Recently, $f(Q)$ gravity has been taken into serious  consideration \cite {Dialektopoulos:2019mtr,Dimakis:2021gby}, in particular, to explain the late-time acceleration and dark energy issues \cite{Atayde:2021pgb,Anagnostopoulos:2021ydo,Frusciante:2021sio,Bajardi:2020fxh,Capozziello:2022wgl} but it is not investigated in details in early universe.

In this paper, we  investigate  $f(Q)$ inflationary cosmology adopting both the potential-slow-roll (PSR) and the Hubble-slow-roll (HSR) mechanisms in view to match observational data. Slow-roll inflation has been widely investigated in several alternative theories of gravity \cite{Vasilis1,Vasilis2,Vasilis3,Vasilis4} because it constitutes a useful approach, in the framework  of inflationary paradigm, to achieve viable early-time models. 

In the PSR frame, we deal with the potentials derived from some  forms of $f(Q)$ function and results will be matched with Planck and BICEP2/Keck array datasets in order to investigate the viability of the $f(Q)$ approach. This means that we are performing a conformal transformation and we study dynamics in the Einstein frame. On the contrary, adopting the HSR viewpoint, we carry out the inflationary analysis for some conventional potentials like  monomial potential (with both integer and fractional powers), exponential potential, and natural inflationary potential for a specific form $f(Q)=\alpha Q+\beta Q^{2}$, proposed in Ref.\cite{BeltranJimenez:2019tme}, which is a suitable prescription for the early time inflation. This means that HSR is considering a Jordan frame where cosmological background is not conformally transformed. 

The layout of the paper is the following.  In Sec.\ref{fq}, we briefly sketch  the main features of $f(Q)$ gravity and cosmology. In particular, we develop the conformal transformation for $f(Q)$ gravity pointing out differences with other extended theories of gravity like $f(R)$.  Sec.\ref{psr} is devoted to the PSR inflation considering some suitable potentials derived from $f(Q)$ gravity in the Einstein frame. The obtained results are compared  with observational datasets coming from the cosmic microwave background (CMB) anisotropies. The aim  is to find  observational constraints on $f(Q)$ parameters space and   their predictions about the spectrum. In Sec.\ref{hsr}, we study the issue in the HSR approach, that is remaining in the Jordan frame. We use some conventional potentials in the context of the model $f(Q)\propto Q^{2}$. Discussion and  conclusions are reported in Sec.\ref{con}. 

\section{$f(Q)$ gravity and cosmology}\label{fq}
In order to define $f(Q)$ gravity, we have to 
 start our considerations with a generic affine connection \begin{equation}
\Gamma^{\alpha}{}_{\mu\nu}=\{^{\alpha}_{\mu\nu}\}+K^{\alpha}{}_{\mu\nu}+L^{\alpha}{}_{\mu\nu},
\label{1}    
\end{equation}
where the Levi-Civita connection, the contorsion tensor and the disformation tensor are defined, respectively as  
\begin{equation}
\{^{\alpha}_{\mu\nu}\}=\frac{1}{2}g^{\alpha\lambda}(g_{\mu\lambda,\nu}+g_{\lambda\nu,\mu}-g_{\mu\nu,\lambda}),
\label{2}    
\end{equation}
\begin{equation}
K^{\alpha}{}_{\mu\nu}=\frac{1}{2}g^{\alpha\lambda}(T_{\mu\lambda\nu}+T_{\nu\lambda\mu}+T_{\lambda\mu\nu}),
\label{3}    
\end{equation}
\begin{equation}
L^{\alpha}{}_{\mu\nu}=\frac{1}{2}g^{\alpha\lambda}(Q_{\lambda\mu\nu}-Q_{\mu\lambda\nu}-Q_{\nu\lambda\mu}).
\label{4}    
\end{equation}
The torsion and non-metricity tensors are introduced as
\begin{equation}
T^{\alpha}{}_{\mu\nu}\equiv\Gamma^{\alpha}{}_{\mu\nu}-\Gamma^{\alpha}{}_{\nu\mu},
\label{5}    
\end{equation}
\begin{equation}
Q_{\alpha\mu\nu}\equiv\nabla_{\alpha}g_{\mu\nu}=\partial_{\alpha}g_{\mu\nu}-\Gamma^{\lambda}{}_{\alpha\mu}g_{\lambda\nu}-\Gamma^{\alpha}{}_{\alpha\nu}g_{\mu\lambda}\,.
\label{6}    
\end{equation}
Using the inverse metric, non-metricity tensor can be expressed as $Q_{\alpha}{}^{\mu\nu}=-\nabla_{\alpha}g^{\mu\nu}$. The curvature Riemann tensor  is defined as
\begin{equation}
R^{\alpha}{}_{\lambda\mu\nu}\equiv2\partial_{[\mu}\Gamma^{\alpha}{}_{\nu]\lambda}+2\Gamma^{\alpha}{}_{[\mu|\beta|}\Gamma^{\beta}{}_{\nu]\lambda}.
\label{7}    
\end{equation}
In GR, the fundamental geometric object is the  curvature  while  torsion and  non-metricity tensors are zero. Consequently, the affine connection  is the Levi-Civita connection (\ref{2}). In TEGR, the Ricci and the non-metricity tensors and scalars are zero and the fundamental object is the torsion scalar $T=S^{\alpha\mu\nu}T_{\alpha\mu\nu}$ where $S^{\alpha\mu\nu}=\frac{1}{2}(K^{\mu\nu\alpha}-g^{\alpha\nu}T^{\lambda\mu}{}_{\lambda}+g^{\alpha\mu}T^{\lambda\nu}{}_{\lambda})$. Then, the affine connection takes the Weitzenb\"{o}ck form $\Gamma^{\alpha}{}_{\mu\nu}=e^{\alpha}_{\lambda}\partial_{\nu}e^{\lambda}_{\mu}$. 

In STEGR, the Ricci and the torsion tensors and scalas are zero and we deal with the non-metricity scalar
\begin{equation}
Q=-Q_{\alpha\mu\nu}P^{\alpha\mu\nu},
\label{8}
\end{equation}
where the non-metricity conjecture is defined as
\begin{equation}
P^{\alpha}{}_{\mu\nu}=-\frac{1}{2}L^{\alpha}{}_{\mu\nu}+\frac{1}{2}(Q^{\alpha}-\hat{Q}^{\alpha})g_{\mu\nu}-\frac{1}{4}\delta^{\alpha}_{(\mu}Q_{\nu)}.
\label{9}
\end{equation}
Here, $Q_{\alpha}=g^{\mu\nu}Q_{\alpha\mu\nu}$ and $\hat{Q}_{\alpha}=g^{\mu\nu}Q_{\mu\alpha\nu}$ are the two independent traces of the non-metricity tensor. The most general connection of STEGR is then
\begin{equation}
\Gamma^{\alpha}{}_{\mu\nu}:=\frac{\partial x^{\alpha}}{\partial \xi^{\lambda}}\frac{\partial^{2}\xi^{\lambda}}{\partial x^{\mu}\partial x^{\nu}},
\label{10}
\end{equation}
where $\xi^{\lambda}=\xi^{\lambda}(x)$ is an arbitrary function of spacetime position. This connection can be obtained from vanishing connections under the transformation $x^{\mu}\rightarrow\xi^{\mu}(x^{\nu})$. Thanks to the coincident gauge, it is always possible to determine a coordinate transformation in order to obtain a  vanishing connection $\Gamma^{\alpha}{}_{\mu\nu}$ and then the non-metricity tensor reduces to $Q_{\alpha\mu\nu}=\partial_{\alpha}g_{\mu\nu}$. 

In STEGR, the fundamental object of gravitational field is derived from   non-metricity scalar and the Hilbert-Einstein  action becomes
\begin{equation}
S=\int{d^{4}x\sqrt{-g}\Big\{\frac{Q}{2}+\mathcal{L}_{m}(g_{\mu\nu},\Psi_{m})\Big\}}.
\label{11}    
\end{equation}
Analogous to  $f(R)$ and $f(T)$  gravity, we  can introduce an extended version of STEGR, where the action is given by
\begin{equation}
S=\int{d^{4}x\sqrt{-g}\Big\{-\frac{f(Q)}{2}+\mathcal{L}_{m}(g_{\mu\nu},\Psi_{m})\Big\}},
\label{12}
\end{equation}
where $g$ is the determinant of the metric $g_{\mu\nu}$ and $\mathcal{L}_{m}$ is the Lagrangian of matter depending on the metric $g_{\mu\nu}$ and matter fields $\Psi_{m}$ filling the universe as a perfect fluid. Here, we suppose $\kappa^{2}\equiv8\pi G=1$. Note that for  $f(Q)=Q$, we recover STEGR as the equivalent version of GR. 

In the following, we consider a spatially flat universe described by the FLRW metric as
\begin{equation}
ds^{2}=-dt^{2}+a(t)^{2}(dx^{2}+dy^{2}+dz^{2}),   
\label{13}
\end{equation}
where $a(t)$ is the scale factor of the universe. Starting from the action \eqref{12}, cosmological  equations are
\cite{BeltranJimenez:2019tme}
\begin{equation}
6f'H^{2}-\frac{1}{2}f=\rho,\hspace{1cm}(12f''H^{2}+f')\dot{H}=-\frac{1}{2}(\rho+p),
\label{14}
\end{equation}
where $'$ and $''$ denote the first and second derivatives with respect to $Q$ and the dot denotes the derivative with respect to the cosmic time. Here, $\rho$ and $p$ are the energy density and pressure of the perfect fluid. The above expressions can be written in the form of  standard Friedman equations 
\begin{equation}
H^{2}=\frac{1}{3}\rho_{eff},\hspace{1cm}2\dot{H}+3H^{2}=-p_{eff},
\label{15}   
\end{equation}
where $\rho_{eff}$   and $p_{eff}$ are the effective density and pressure of the total fluid  defined as 
\begin{equation}
\rho_{eff}=\frac{\rho+\frac{f}{2}}{2f'},\hspace{1cm}p_{eff}=\frac{p+3H^{2}(f'+8\dot{H}f'')-\frac{f}{2}}{f'}. 
\label{16}
\end{equation}
Now, we can define the effective equation of state (EoS) as
\begin{equation}
w_{eff}=\frac{p_{eff}}{\rho_{eff}}=\frac{4p+12H^{2}(f'+8\dot{H}f'')-2f}{2\rho+f}\,,
\label{17}    
\end{equation}
and then $f(Q)$ gravity can potentially addresses DE issues thanks to the value of $w_{eff}$ \cite{Frusciante:2021sio, Anagnostopoulos:2021ydo}.

In this framework, it is possible to investigate conformal transformations for $f(Q)$ gravity in order to compare similar features to $f(R)$ and $f(T)$ gravity. In general, we have the conformal transformation
\begin{equation}
\tilde{g}_{\mu\nu}=\Omega^{2}g_{\mu\nu}\,,
\label{18}
\end{equation}
 where the conformal factor $\Omega=\Omega(\varphi(x))$ is a differentiable and non-singular function. It allows to transform dynamics from the Jordan to the Einstein frame and vice-versa \cite{Marino}. Here, a tilde represents quantities in the Einstein frame. Under the such transformation (\ref{18}), it is possible to define  the following identities
\begin{equation}
\tilde{g^{\mu\nu}}=\Omega^{-2}g^{\mu\nu},\hspace{0.75cm}\sqrt{-\tilde{g}}=\Omega^{4}\sqrt{-g}.
\label{19}    
\end{equation}
For scalar functions, we have $\nabla_{\mu}\Omega=\partial_{\mu}\Omega$, and so $\tilde{\nabla}_{\mu}\Omega=\nabla_{\mu}\Omega$. Recall that $\tilde{\partial_{\mu}}=\partial_{\mu}$ since $x^{\mu}$ is unaffected under the conformal transformation. Using the above relations, the conformal version of non-metricity tensor and non-metricity conjecture in the coincident gauge ($Q_{\alpha\mu\nu}=\nabla_{\alpha}g_{\mu\nu}=\partial_{\alpha}g_{\mu\nu}$) are
\begin{equation}
\tilde{Q}_{\alpha\mu\nu}=\Omega^{2}\Big(Q_{\alpha\mu\nu}+2g_{\mu\nu}\partial_{\alpha}\ln{\Omega}\Big),
\label{20}    
\end{equation}
and
\begin{equation}
\tilde{P}^{\alpha\mu\nu}=-\Omega^{-4}\bigg[P^{\alpha\mu\nu}+\Omega^{2}\partial_{\beta}\ln{\Omega}\Big(\frac{1-n}{2}g^{\alpha\beta}g^{\mu\nu}-g^{(\mu\beta}g^{\nu)\alpha}+\frac{3n}{4}g^{\mu\beta}g^{\alpha\nu}+\frac{n}{4}g^{\nu\beta}g^{\alpha\mu}\Big)\bigg],
\label{21}    
\end{equation}
where the relations $g_{\mu\nu}g^{\mu\nu}=n$, $g_{\mu\nu}g^{\mu\alpha}=\delta_{\nu}^{\alpha}$ have been used.  Here $n\geq2$ is the dimension of  spacetime. For $n=4$, the non-metricity scalar in the Einstein frame $\tilde{Q}=-\tilde{Q}_{\alpha\mu\nu}\tilde{P}^{\alpha\mu\nu}$ takes the form
\begin{equation}
\tilde{Q}=-\Omega^{-2}Q+20g^{\mu\nu}\partial_{\nu}w\partial_{\mu}w+\frac{7}{2}(\Omega^{-2}-1)Q^{\mu}\partial_{\mu}w,
\label{22}
\end{equation}
where $w\equiv\ln\Omega$. The inverse transformation of the relation (\ref{22}) is given by
\begin{equation}
Q=-\Omega^{2}\tilde{Q}+20\Omega^{2}\tilde{g}^{\mu\nu}\partial_{\nu}w\partial_{\mu}w+\frac{7}{2}(1-\Omega^{2})\tilde{Q}^{\mu}\partial_{\mu}w.
\label{23}    
\end{equation}
To obtain the action in the Einstein frame, we first rewrite the action  (\ref{12}) in the Jordan frame as
\begin{equation}
S_{E}=\int{d^{4}x\sqrt{-g}\Big(-\frac{FQ}{2}+U\Big)}+\int{d^{4}x\mathcal{L}_{m}(g_{\mu\nu},\Psi_{M})},\hspace{1cm}with\hspace{1cm}U=\frac{FQ-f}{2}.
\label{24}    
\end{equation}
Using the relation (\ref{23}) and $\sqrt{-g}=\Omega^{-4}\sqrt{-\tilde{g}}$, the above action can be rewritten as
\begin{equation}
S_{E}=\int{d^{4}x\sqrt{-\tilde{g}}\bigg(\frac{1}{2}F\Omega^{-2}\Big[\tilde{Q}-20\tilde{g}^{\mu\nu}\partial_{\nu}w\partial_{\mu}w-\frac{7}{2}\frac{(1-\Omega^{2})}{\Omega^{2}}\tilde{Q}^{\mu}\partial_{\mu}w\Big]+\Omega^{-4}U\bigg)}+\int{d^{4}x\mathcal{L}_{m}(\Omega^{-2}\tilde{g}_{\mu\nu},\Psi_{M})}.
\label{25}    
\end{equation}
Then, by choosing $F=\Omega^{2}$, we have
\begin{equation}
S_{E}=\frac{1}{2}\int{d^{4}x\sqrt{-\tilde{g}}\bigg(\tilde{Q}-20\tilde{g}^{\mu\nu}\partial_{\nu}w\partial_{\mu}w-\frac{7}{2}\frac{(1-F)}{F}\tilde{Q}^{\mu}\partial_{\mu}w+2F^{-2}U\bigg)}+\int{d^{4}x\mathcal{L}_{m}(F^{-1}\tilde{g}_{\mu\nu},\Psi_{M})}.
\label{26}    
\end{equation}
It is worth noticing the presence of the third term in the action \eqref{26}. Comparing with action \eqref{11}, there is an additional scalar-non-metricity coupling   which cannot be removed by a standard conformal transformation. In other words, $f(Q)$ theories are not dynamically equivalent to the STEGR action, involving just $Q$ plus a scalar field via a conformal transformation.  This conformal structure differs from the $f(R)$ case and the situation is similar to that outlined in \cite{Yang:2010ji} for $f(T)$ gravity. However, as discussed in \cite{Yang:2010ji} for $f(T)$,  the further coupling can be neglected in particular regimes.   It is easy to see that, being $\partial_{\mu}w=\partial_{\mu} \ln F= \partial_{\mu} F/F$,  then $(1-F)/F^{3/2}\rightarrow 0$ for $F\rightarrow \infty$. In  this regime also $\ln F$ is slowly varying and then also $\partial_\mu w$ is negligible. In other words,  being $F$  large in strong field regime, the third term in \eqref{26} can be neglected. This is an important feature because the transition between the strong   to the weak field regime can be related to the behavior of the function $F$, that is the conformal factor $\Omega^2$. We are in the strong field for $F\rightarrow\infty$ while we are in the weak field for  $F\rightarrow 0$. It means that conformal transformations work in strong field while, in the weak field, there is a breaking of conformal structure  which is led by  the third term in \eqref{26}. In other words, the behavior of function $F$, in the framework of non-metric gravity,  constitutes a natural mechanism to exit from inflation.

With these considerations in mind and  in strong field regime, if we introduce a new scalar field $\varphi\equiv\sqrt{5}\ln F$, the action in the Einstein frame takes the familiar form
\begin{equation}
S_{E}=\int{d^{4}x\sqrt{-\tilde{g}}\bigg(\frac{1}{2}\tilde{Q}-\frac{1}{2}\tilde{g}^{\mu\nu}\partial_{\mu}\varphi\partial_{\nu}\varphi-V(\varphi)\bigg)}+\int{d^{4}x\mathcal{L}_{m}(F^{-1}\tilde{g}_{\mu\nu},\Psi_{m})},\hspace{0.15cm}
\label{27}    
\end{equation}
where the  potential is 
\begin{equation}
V(\varphi)=\frac{f-FQ}{2F^{2}},\hspace{1cm}\mbox{with}\hspace{1cm}F(Q)\equiv f'(Q)\equiv e^{\sqrt{\frac{1}{5}}\varphi}.
\label{28}    
\end{equation}
We can develop now our analysis of inflation. In particular, two approaches are possible. In the Einstein frame, the potential derived from the function  $f(Q)$ gives the inflaton dynamics. In the second, we fix the $f(Q)$ background with a given inflaton field and study the evolution of the Hubble parameter. In this case, we are in the Jordan frame.

\begin{figure*}[!hbtp]
	\centering
	\includegraphics[width=.55\textwidth,keepaspectratio]{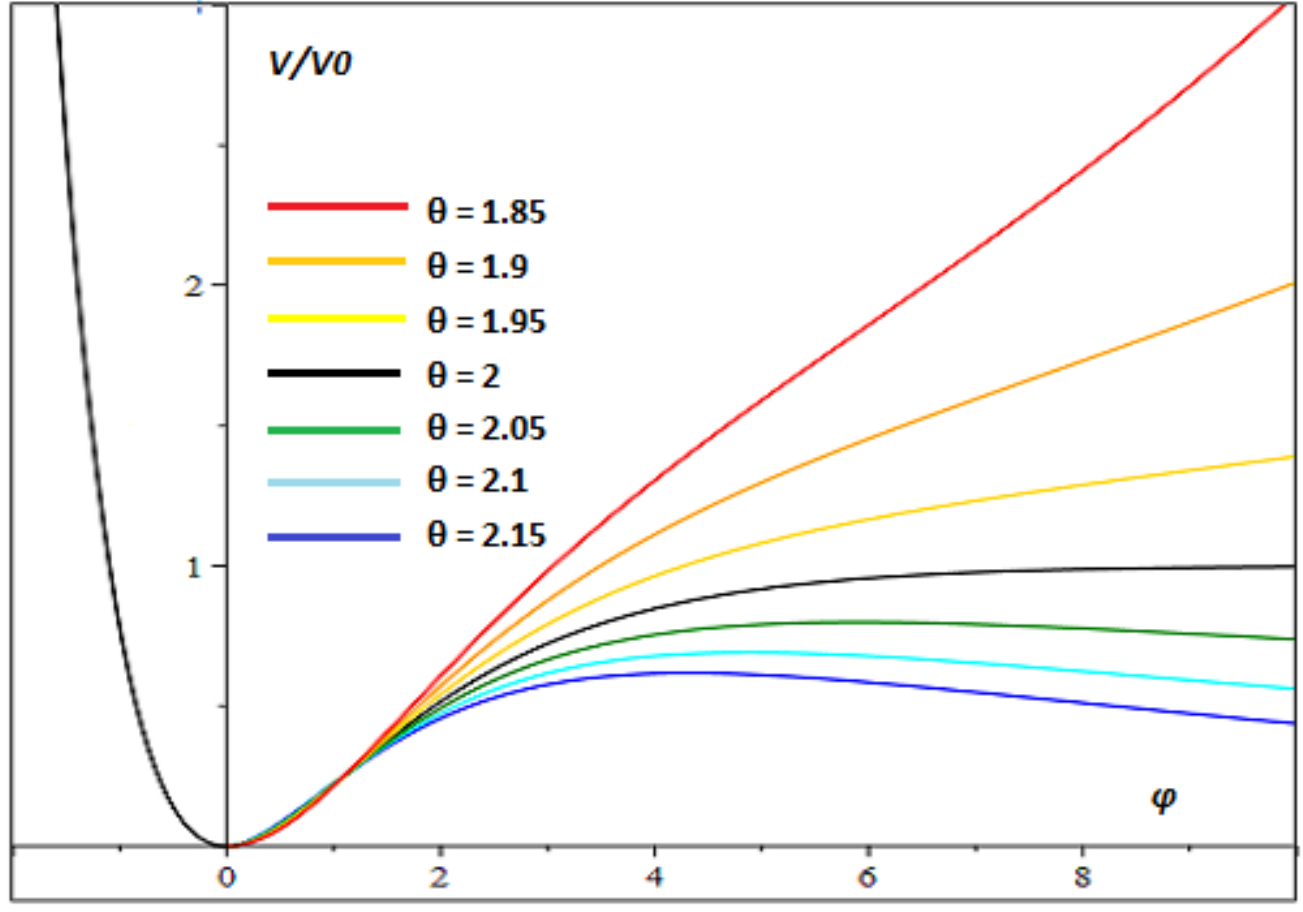}
	\caption{The potential of $Q^{\theta}$ model (\ref{29}) for different values of $\theta$. For $\theta=2$, we reproduce the result of the $Q^{2}$ model suggested in \cite{BeltranJimenez:2019tme} for high energy regimes. For $\theta<2$, the potential ascends towards the higher values of $\varphi$ while it decreases to zero steeper than $\theta=2$. For $\theta>2$, the potential shows a maximum $\varphi_{m}$ and then it tends to zero for large values of $\varphi$.}
	\label{fignew1}
\end{figure*}
\section{Potential-slow-roll inflation}\label{psr}
In this section, we shall study the PSR approach to inflation using the potentials corresponding to different forms of $f(Q)$ gravity. The potentials are derived after a conformal transformation and we study dynamics in the Einstein frame adopting a strong-field regime. In the slow-roll approximation is also easy to see that the coupled kinetic term in action \eqref{26}, the third term,  is negligible.
\subsection{The $Q^{\theta}$ model}
Let us first study the $Q^{\theta}$ model defined as
\begin{equation}
f(Q)=Q+\xi Q^{\theta},
\label{29}
\end{equation}
where the parameter $\theta>1$ is a real number responsible for the accelerating phase in the early universe. See also \cite{BeltranJimenez:2019tme}. Here, $\xi$ and $\theta$  are the free parameters of the model. Adapting  the conformal transformation (\ref{18}), we obtain the potential corresponding to the above $f(Q)$, that is 
\begin{equation}
V(\varphi)=V_{0}e^{-2\sqrt{\frac{1}{5}}\varphi}(e^{\sqrt{\frac{1}{5}}\varphi}-1)^{\frac{\theta}{\theta-1}},
\label{30}
\end{equation}
where $V_{0}=\frac{1}{2}(1-\theta)\theta^{\frac{\theta}{1-\theta}}\xi^{\frac{1}{1-\theta}}$. Fig.\ref{fignew1} presents the behaviour of the potential (\ref{30}) for different values of $\theta$. Let us review some important properties of the obtained potential. For $\theta=2$, we recover the potential of $f(Q)$ model proposed in \cite{BeltranJimenez:2019tme} which is relevant in the high curvatures regime, in particular, for the inflationary era. In such a case, the potential (\ref{30}) is given by $V(\varphi)=-\frac{1}{8\xi}(1-e^{-\sqrt{\frac{1}{5}}\varphi})^{2}$. For $\theta<2$, the potential increases gradually but its decreasing towards zero is  steeper than $\theta=2$ model. For $\theta>2$, the potential shows a maximum in $\varphi_m =\sqrt{\frac{1}{5}}\ln{\frac{2(\theta-1)}{\theta-2}}$ so that it is approaching to zero for large value of $\varphi$. Inflation occurs for $0 \leq \varphi \leq \varphi_m$ and $\varphi > \varphi_m$. 

The slow-roll parameters in the PSR formalism are defined as \begin{equation}
\epsilon=\frac{1}{2}\bigg(\frac{V'(\varphi)}{V(\varphi)}\bigg)^{2},\qquad\quad \eta=\frac{V''(\varphi)}{V(\varphi)},
\label{31}
\end{equation}
where prime denotes derivative with respect to the scalar field $\varphi$. Also, the number of e-folds $N$ is given by
\begin{equation}
N\equiv\frac{1}{\sqrt{2}}\int^{\varphi_{i}}_{\varphi_{e}}{\frac{V}{V'}d\varphi} \equiv\int^{\varphi_{i}}_{\varphi_{e}}{\frac{1}{\sqrt{2\epsilon}}d\varphi},
\label{32}
\end{equation}
where subscribes $i$ and $e$ denote the start and end of inflation, respectively. Note that during inflation $\epsilon\ll1$ and $\eta\ll1$ while inflation ends when either $\epsilon=1$ or $\eta=1$ is fulfilled. The spectral parameters, \textit{i.e.} the spectral index and the tensor-to-scalar ratio are expressed as
\begin{equation}
n_{s}=1-6\epsilon+2\eta,\quad\quad\quad r=16\epsilon.
\label{33}  
\end{equation}
In the following, we study the model for two cases $\theta\neq2$ and $\theta=2$, separately. 
\begin{figure*}[!hbtp]
	\centering
	\includegraphics[width=.70\textwidth,keepaspectratio]{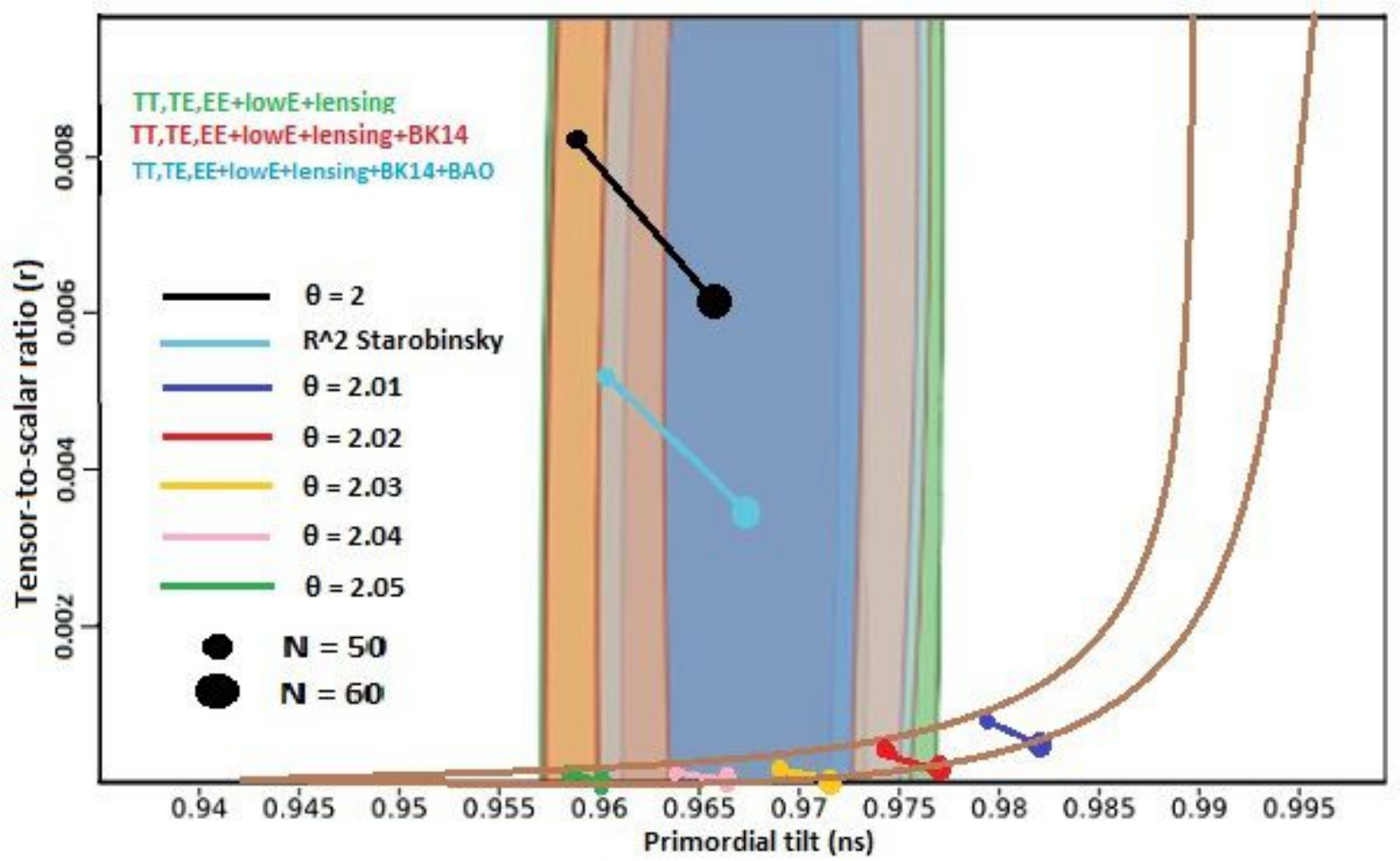}
	\caption{The marginalized joint 68\% and 95\% CL regions for $n_{s}$ and $r$ at $k = 0.002$ Mpc$^{-1}$ from Planck alone and in combination with BK14 or BK14+BAO data \cite{cmb} and the $n_{s}-r$ constraints on the $Q^{\theta}$ model (\ref{29}) for two cases $\theta=2$ and $\theta\neq2$ compared with the $R^{2}$ Starobinsky model.}
	\label{fignew2}
\end{figure*}
\subsubsection{The case $\theta\neq2$}
Plugging the potential (\ref{30}) into Eqs. (\ref{31}), the slow-roll parameters of the case $\theta\neq2$ are 
\begin{equation}
\epsilon=\frac{\big((\theta-2)e^{\sqrt{\frac{1}{5}}\varphi}-2(\theta-1)\big)^{2}}{10(\theta-1)^{2}\big(e^{\sqrt{\frac{1}{5}}\varphi}-1\big)^{2}},\hspace{1cm}\eta=\frac{(-5\theta^{2}+13\theta-8)e^{\sqrt{\frac{1}{5}}\varphi}+(\theta-2)^{2}e^{2\sqrt{\frac{1}{5}}\varphi}+4(\theta -1)^{2}}{5(\theta-1)^{2}\big(e^{\sqrt{\frac{1}{5}}\varphi}-1\big)^{2}}.
\label{34}
\end{equation}
By setting $\epsilon=1$, we find the value of $\varphi$ at the end of inflation as
\begin{equation}
\varphi_{end}=\sqrt{5}\ln\Big(\frac{(\theta-1)(\sqrt{10}-2)}{\sqrt{10}(\theta-1)-(\theta-2)}\Big).
\label{35}
\end{equation}
The number of e-folds (\ref{32}) of the model takes the form
\begin{equation}
N\simeq-\frac{5\theta}{2(\theta-2)}\ln\bigg(\frac{(\theta-2)e^{\sqrt{\frac{1}{5}}\varphi_{i}}}{2(1-\theta)}+1\bigg).
\label{36}
\end{equation}
The spectral parameters (\ref{33}) of the model are
\begin{equation}
n_{s}\simeq\frac{4\theta(-4\theta+3)e^{-\frac{2(\theta-2)N}{5\theta}}+(16\theta^{2}-24\theta+4)e^{-\frac{4(\theta-2)N}{5\theta}}+\theta^{2}+8\theta}{5\Big((2\theta-2)e^{-\frac{2(\theta-2)N}{5\theta}}-\theta\Big)^2},\hspace{1cm}r\simeq\frac{32(\theta-2)^{2}e^{-\frac{4(\theta-2)N}{5\theta}}}{5\Big((2\theta-2)e^{-\frac{2(\theta-2)N}{5\theta}}-\theta\Big)^2}.
\label{37}    
\end{equation}
In Fig.\ref{fignew2}, the comparison with data is presented.
\subsubsection{The case $\theta=2$}
In the specific case $\theta=2$, the potential (\ref{30}) takes the following form
\begin{equation}
V(\varphi)=V_{0}(1-e^{-\sqrt{\frac{1}{5}}\varphi})^2,
\label{38}    
\end{equation}
where ${\displaystyle V_{0}=-\frac{1}{8\xi}}$. 
This case is particularly interesting because it can be compared directly to the $R^2$ Starobinsky model.

Using the definitions (\ref{31}), we have the slow-roll parameters  as
\begin{equation}
\epsilon=\frac{2}{5\big(e^{\sqrt{\frac{1}{5}}\varphi}-1\big)^{2}},\hspace{1cm}\eta=\frac{2\big(2-e^{\sqrt{\frac{1}{5}}\varphi}\big)}{5\big(e^{\sqrt{\frac{1}{5}}\varphi}-1\big)^{2}}.
\label{39}    
\end{equation}
Applying the condition $\epsilon=1$ for ending inflation, we have
\begin{equation}
\varphi_{end}=\sqrt{5}\ln\Big(1+\sqrt{\frac{2}{5}}\Big).
\label{40}
\end{equation}
Also, the number of e-folds (\ref{32}) in this case is given by
\begin{equation}
N\simeq\frac{5}{2}e^{\sqrt{\frac{1}{5}}\varphi_{i}}.
\label{41}    
\end{equation}
The spectral parameters (\ref{33})  are, in this case,
\begin{equation}
n_{s}\simeq\frac{4N^{2} -28N+5}{(2N-5)^{2}},\hspace{1cm}r\simeq\frac{160}{(2N-5)^{2}}.
\label{42}    
\end{equation}
Fig.\ref{fignew2} discloses the $n_{s}-r$ constraints coming from the marginalized joint 68\% and 95\% CL regions of the Planck 2018 data in combination with BK14+BAO data on the $Q^{\theta}$ model (\ref{29}) for the two cases $\theta=2$ and $\theta\neq2$. Also, the predication of the $R^{2}$ Starobinsky model as a well-know inflationary model of $f(R)$ gravity is presented. By taking a look at the plot, one can find that  the $R^{2}$ model and the $\theta=2$ model show the match with observational values of $n_{s}$ and $r$ at both 68\% and 95\% C.L. for all three observational datasets. By considering the Planck alone dataset, the figure tells us that the powers $2.02\leq\theta\leq2.05$ present the observational  values of $n_{s}$ and $r$ at the 68\% C.L. while it reduces to $2.03\leq\theta<2.05$ at the 95\% C.L. By combination of the Planck and the BK14, we realize that the observational constraints are similar with the Planck alone case. For a full consideration of the CMB anisotropy observations \textit{i.e.} the Planck+BK14+BAO datesets, we find that the obtained values of $n_{s}$ and $r$ for the powers $2.02\leq\theta\leq2.05$ and $2.03\leq\theta\leq2.04$ are in good agreement with the observations at the 68\% and 95\% C.L., respectively. 

In addition to the above results, the figure shows that the $Q^{\theta}$ model for $\theta\neq2$ offers a smaller value of the tensor-to-scalar ratio $r$ in comparison with the $R^{2}$ Starobinsky model while for $\theta=2$ it shows a bigger value of $r$ than the Starobinsky model. Moreover, large values of $\theta$ are excluded since the observationally allowed deviation of $\theta$ from 2 is very small ($\sim10^{-2}$).

\subsection{Logarithmic corrected model}
Another interesting case is the $f(Q)$ quadratic model
corrected by a  logarithmic term, that is 
\begin{equation}
f(Q)=Q+\xi Q^{2}+\upsilon Q^{2}\ln Q,
\label{43}
\end{equation}
with the phenomenological parameters $\xi$, $\upsilon$. The potential (\ref{28}) of the scalar field $\varphi$,  related to the logarithmic corrected form of $f(Q)$, is
\begin{equation}
V(\varphi)=-\frac{(\xi+\upsilon)Q^{2}(1+\frac{\upsilon}{\xi+\upsilon}\ln Q)}{2\Big(1+(2\xi+\upsilon)Q(1+\frac{2\upsilon}{2\xi+\upsilon}\ln Q)\Big)^{2}}.
\label{44}    
\end{equation}
Here, we recover the $\theta=2$ model in the limit $\upsilon\rightarrow0$. Using the definition of $F$ (\ref{28}), we find $Q$ in terms of the Lambert function $W_{l}$ as
\begin{equation}
Q=\frac{e^{\sqrt{\frac{1}{5}}\varphi}-1}{2\upsilon W_{l}(X)},\hspace{1cm}\mbox{with}\hspace{1cm}X\equiv\frac{e^{\sqrt{\frac{1}{5}}\varphi}-1}{2\upsilon}e^{\frac{2\xi+\upsilon}{2\upsilon}},
\label{45}  
\end{equation}
\begin{figure*}[!hbtp]
	\centering
	\includegraphics[width=.55\textwidth,keepaspectratio]{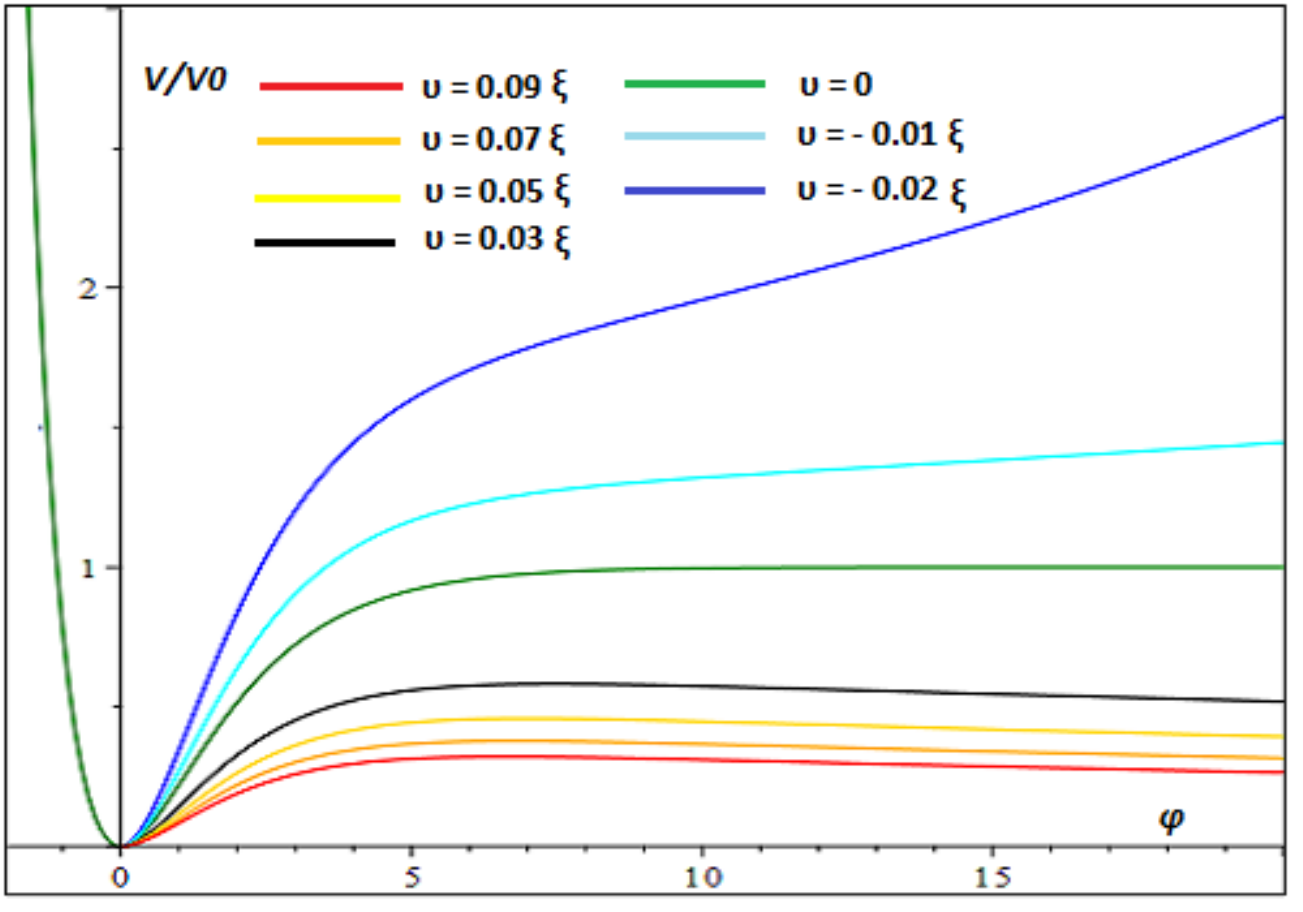}
	\caption{The potential of the logarithmic corrected model (\ref{43}) for different values of $\upsilon$ and $\xi\sim\mathcal{O}(10^{-8})$. The case $\upsilon=0$ recovers the well-known function $f\propto Q^{2}$ \cite{BeltranJimenez:2019tme} suitable for the inflationary era. For $\upsilon<0$, the potential goes upward so that it shows a lager value of $r$ because of a larger slope. For $\upsilon>0$, the potential proposes a maximum value and then it presents a decreasing behaviour toward zero for bigger values of $\varphi$.}
	\label{fignew3}
\end{figure*}
where $W_{l}$ is the Lambert function of branch $l = 0$ for $\upsilon>0$, and , $l = -1$ when $\upsilon < 0$. Using the iterative method and working with just the leading order in $Q$, we find the potential of the model for $\upsilon\ll\xi$ as
\begin{equation}
V(\varphi)\simeq V_{0}\frac{(1-e^{-\sqrt{\frac{1}{5}}\varphi})^{2}}{1+\frac{\upsilon}{2\xi}+\frac{\upsilon}{\xi}\ln\Big(\frac{e^{\sqrt{\frac{1}{5}}\varphi}-1}{2\xi}\Big)},
\label{46}  
\end{equation}
where $V_{0}=-\frac{1}{8\xi}$. In Fig.\ref{fignew3}, we present the behaviour of the obtained potential (\ref{46}) for different values of $\upsilon$ and $\xi\sim\mathcal{O}(10^{-8})$ in which $\upsilon=0$ recovers the case of $\theta=2$ \cite{BeltranJimenez:2019tme}. For $\upsilon<0$, inflationary tilts upward and it shows larger values of the scalar-to-tensor ratio $r$ due to a larger slope. While for $\upsilon>0$, inflation has a maximum value in unstable point and then it runs away for bigger values of $\varphi$.

The slow-roll parameters (\ref{31}) of the model are
\begin{equation}
\epsilon=\frac{2\Big(2\upsilon\ln(\frac{e^{\sqrt{\frac{1}{5}}\varphi}-1}{2\xi})+(1-e^{\sqrt{\frac{1}{5}}\varphi})\upsilon+2\xi\Big)^{2}}{5\Big(2\upsilon\ln(\frac{e^{\sqrt{\frac{1}{5}}\varphi}-1}{2\xi})+\upsilon+2\xi\Big)^{2}\big(e^{\sqrt{\frac{1}{5}}\varphi}-1\big)^{2}},
\label{47}    
\end{equation}
\begin{eqnarray}
&\!&\!\eta=\frac{1}{5\Big(2\upsilon\ln(\frac{e^{\sqrt{\frac{1}{5}}\varphi}-1}{2\xi})+\upsilon+2\xi\Big)^{2}\big(e^{\sqrt{\frac{1}{5}}\varphi}-1\big)^{2}}\Bigg\{-8\upsilon^{2}(e^{\sqrt{\frac{1}{5}}\varphi}-2)\ln\Big(\frac{e^{\sqrt{\frac{1}{5}}\varphi}-1}{\xi}\Big)^{2}-4\Big((-4\upsilon\ln({2})+4\xi+5\upsilon)\times
\nonumber\\&\!&\!
\times e^{\sqrt{\frac{1}{5}}\varphi}+8\upsilon\ln({2})-8\xi-4\upsilon\Big)\upsilon\ln\big(\frac{e^{\sqrt{\frac{1}{5}}\varphi}-1}{\xi}\big)-4\big(-2\upsilon\ln(2)+2\xi+\upsilon\big)\big(-\upsilon\ln(2)+\xi+2\upsilon\big)e^{\sqrt{\frac{1}{5}}\varphi}+8e^{2\sqrt{\frac{1}{5}}\varphi}\upsilon^{2}+\nonumber\\&\!&\!
+4\big(-2\upsilon\ln(2)+2\xi+\upsilon\big)^{2}\Bigg\}.
\label{48}    
\end{eqnarray}
\begin{figure*}[!hbtp]
	\centering
	\includegraphics[width=.70\textwidth,keepaspectratio]{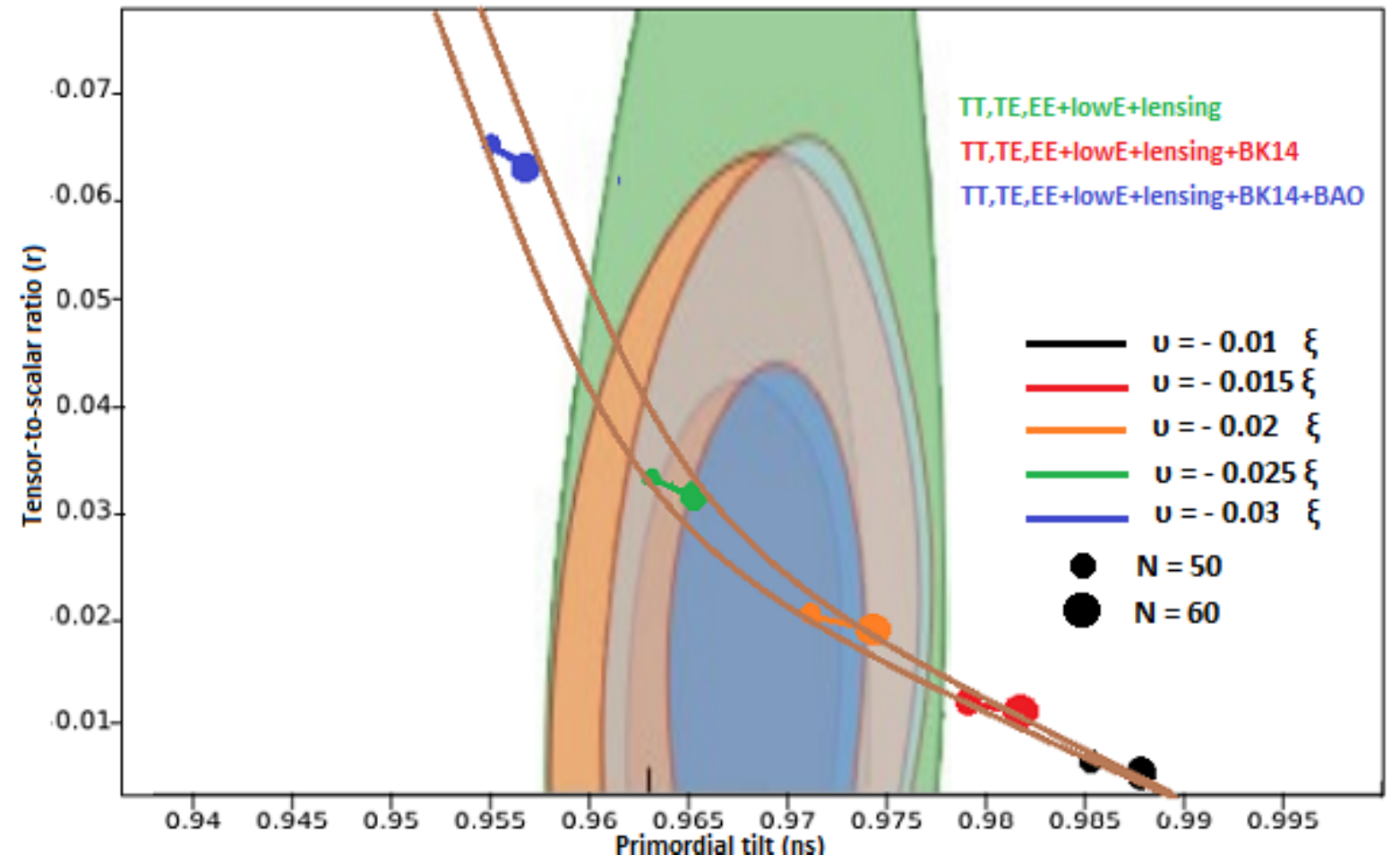}
	\caption{The marginalized joint 68\% and 95\% CL regions for $n_{s}$ and $r$ at $k = 0.002$ Mpc$^{-1}$ from Planck alone and in combination with BK14 or BK14+BAO data \cite{cmb} and the $n_{s}-r$ constraints on the logarithmic corrected model (\ref{43}) for different values of $\upsilon$ and $\xi\sim\mathcal{O}(10^{-8})$.}
	\label{fignew4}
\end{figure*}
The value of the scalar field $\varphi$ at the end of inflation can be obtained from
\begin{equation}
1=\frac{2\Big(2\upsilon\ln(\frac{e^{\sqrt{\frac{1}{5}}\varphi_{end}}-1}{2\xi})+(1-e^{\sqrt{\frac{1}{5}}\varphi_{end}})\upsilon+2\xi\Big)^{2}}{5\Big(2\upsilon\ln(\frac{e^{\sqrt{\frac{1}{5}}\varphi_{end}}-1}{2\xi})+\upsilon+2\xi\Big)^{2}\big(e^{\sqrt{\frac{1}{5}}\varphi_{end}}-1\big)^{2}}.
\label{49}    
\end{equation}
The number of e-folds (\ref{32}) for the logarithmic corrected model is 
\begin{equation}
N\simeq-\frac{3\ln\big(\frac{1-\frac{\tau^{2}}{3\hat{\epsilon}}}{1-\frac{\tau^{2}}{3}}\big)-6\tanh^{-1}(\frac{\tau}{\sqrt{3\hat{\epsilon}}})}{\tau(2+\tau)},
\label{50}    
\end{equation}
where $\tau\equiv\frac{\upsilon\Big(1+\frac{\upsilon}{2\xi}+\frac{\upsilon}{\xi}\ln\big(\frac{e^{\sqrt{\frac{1}{5}}\varphi}-1}{2\xi}\big)\Big)^{-1}}{\xi}$ and $\hat{\epsilon}$ is the first slow-roll parameter of the case $\theta=2$ (\ref{39}). Now, from eq.(\ref{33}) and in the limit of $\upsilon\ll\xi$, the
inflationary parameters are  
\begin{eqnarray}
&\!&\!n_{s}\simeq1-\frac{12\Big(\frac{2\xi(6-2N)}{3N}+2\upsilon(\ln 2-\mathcal{G})-\upsilon-2\xi\Big)^{2}}{5\big(2\upsilon(\ln 2-\mathcal{G})-\upsilon-2\xi\big)^{2}\Big(\frac{2\xi(6-2N)}{3N\upsilon}-1\Big)^{2}}+\frac{1}{10\Big(\frac{2\xi(6-2N)}{3N\upsilon}-1\Big)^{2}\Big(-\upsilon\ln 2+\upsilon \mathcal{G}+\xi+\frac{\upsilon}{2}\Big)^{2}}\times\nonumber\\&\!&\!
\times\Bigg\{-8\upsilon^{2}\mathcal{G}^{2}\Big(\frac{2\xi(6-2N)}{3N\upsilon}-2\Big)^{2}-16\upsilon\mathcal{G}\Big(\frac{2\xi(6- 2N)(-\upsilon\ln 2+\xi+\frac{5}{4\upsilon})}{3N\upsilon}+2\upsilon\ln 2-2\xi-\upsilon\Big)-\frac{16\xi(6-2N)}{3N\upsilon}\times\nonumber\\&\!&\!
\times\Big(-\upsilon\ln 2+\xi+ \frac{1}{2}\upsilon\Big)\big(-\upsilon\ln 2+2\upsilon+\xi\big)+\frac{32\xi^{2}(6-2N)^{2}}{9N^{2}}+ 16\big(-\upsilon\ln 2+\xi+\frac{\upsilon}{2}\big)^{2}\Bigg\},
\label{51}    
\end{eqnarray}
\begin{equation}
r\simeq\frac{32\Big(\frac{2\xi(6-2N)}{3N}+2\upsilon(\ln 2-\mathcal{G})-\upsilon-2\xi\Big)^{2}}{5\big(2\upsilon(\ln 2-\mathcal{G})-\upsilon-2\xi\big)^{2}\Big(\frac{2\xi(6-2N)}{3N\upsilon}-1\Big)^{2}},
\label{52}    
\end{equation}
where $\mathcal{G}\equiv\ln\Big(\frac{2\xi(6 -2N)-3N\upsilon}{3N\upsilon\xi}\Big)$. 

In Fig.\ref{fignew4}, we draw the $n_{s}-r$ constraints coming from the marginalized joint 68\% and 95\% CL regions of the Planck 2018 data in combination with BK14+BAO data on the logarithmic corrected model (\ref{43}) for different values of $\upsilon$ and $\xi\sim\mathcal{O}(10^{-8})$. From the Planck alone and its combination with the BK14, we find the observational constrain $-0.03\xi<\upsilon<-0.015\xi$ at the 68\% C.L. while it turns to $-0.025\xi\leq\upsilon\leq-0.02\xi$ at the 95\% C.L. A full combination of the Planck with the BK14 and BAO datasets discloses that the obtained values of $n_{s}$ and $r$ are related to the intervals $-0.025\xi\leq\upsilon<-0.015\xi$ and $-0.025\xi\leq\upsilon\leq-0.02\xi$ at the 68\% C.L. and 95\% C.L., respectively.   
\section{Hubble-slow-roll inflation}\label{hsr}
The above analysis can be developed in the Jordan frame without conformally transforming the $f(Q)$ function. To this aim, we consider
 the HSR approach  assuming  inflationary models for a   $f(Q)$ function of the form 
\begin{equation}
f(Q)=\alpha Q+\beta Q^{m}\,,
\label{53}    
\end{equation}
where $\alpha$ and $\beta$ are the parameters of the model. Also, $m$ is a dimensionless parameter so that $m<1$ corresponds to the low-curvature regime (suitable for DE) and $m>1$ is relevant to the high curvature regime (suitable for inflation) \cite{BeltranJimenez:2019tme}. In the following, we build our analysis on the case of $m=2$. Being $Q=6H^{2}$, \cite{Capozziello:2022wgl}, one can rewrite the dynamical equations (\ref{14}) as
\begin{equation}
\alpha H^{2}+17\beta H^{4}=\frac{1}{3}\rho,\hspace{1cm}\alpha\dot{H}+36\beta H^{2}\dot{H}=-\frac{1}{2}(\rho+p),
\label{54}
\end{equation}
where dot refers to the derivative with respect to the cosmic time $t$. Regarding the  inflationary mechanism, the universe can be considered as filled by a single scalar field acting as the source, that is the \textit{inflaton}, with the energy density and pressure
\begin{equation}
\rho_{\varphi}=\frac{\dot{\varphi}^{2}}{2}+V(\varphi),\hspace{1cm}p_{\varphi}=\frac{\dot{\varphi}^{2}}{2}-V(\varphi),
\label{55}
\end{equation}
where $V$ is the potential of the inflaton field. Plugging the above relations into Eqs. (\ref{54})  under the slow-roll approximation $\ddot{\varphi}\ll H\dot{\varphi}$ and $\frac{\dot{\varphi}^{2}}{2}\ll V(\varphi)$, we have
\begin{equation}
H^{2}=\frac{-3\alpha\pm\sqrt{9\alpha^{2}+4\gamma V}}{2\gamma}
\label{56},
\end{equation}
where $\gamma=54\beta$.  In the following we will work with $\gamma$ instead of $\beta$. Choosing the solution with the positive sign, the slow-roll parameters of the model are calculated as
\begin{equation}
\epsilon=-\frac{\dot{H}}{H^{2}}=-\frac{\sqrt{2\gamma^{3}}V'\dot{\varphi}}{\sqrt{(9\alpha^{2}+4\gamma V)(-3\alpha+\sqrt{9\alpha^{2}+4\gamma V})^{3}}},
\label{57}
\end{equation}
\begin{eqnarray}
&\!&\!\eta=-\frac{\ddot{H}}{2H\dot{H}}=-\frac{\sqrt{2\gamma }}{2V'\dot{\varphi}(9\alpha^{2}+4\gamma V)\sqrt{\Big(-3\alpha+\sqrt{9\alpha^{2}+4\gamma V}\Big)^{3}}}\Bigg\{\bigg(-(9\alpha^{2}+4\gamma V)V''\Big(3\alpha-\sqrt{9\alpha^{2}+4\gamma V}\Big)+\nonumber\\&\!&\!
+3\gamma V'^{2}\Big(2\alpha-\sqrt{9\alpha^{2}+4\gamma V}\Big)\bigg)\dot{\varphi}^{2}-\ddot{\varphi}V'(9\alpha^{2}+4\gamma V)\Big(3\alpha-\sqrt{9\alpha^{2}+4\gamma V}\Big)\Bigg\},
\label{58}
\end{eqnarray}
where the prime denotes the derivative with respect to the scalar field $\varphi$. During the inflationary era, $\epsilon\ll1$ and inflation ends when $\epsilon=1$ or $\eta=1$. The number of e-folds of the model is given by
\begin{equation}
N=\int^{t_{e}}_{t_{i}}{Hdt}=\int^{\varphi_{e}}_{\varphi_{i}}{\sqrt{\frac{-3\alpha+\sqrt{9\alpha^{2}+4\gamma V}}{2\gamma\dot{\varphi}^{2}}}d\varphi},
\label{59}
\end{equation}where $\varphi_{i}$ and $\varphi_{e}$ are the values of inflaton at the beginning   and at  the end of inflation. Also, the spectral parameters, \textit{i.e.} the spectral index and the tensor-to-scalar ratio are defined as in (\ref{33}). 

Now, let us specify the study for some standard forms of inflationary potential \textit{e.g.} monomial, exponential and natural inflationary potentials. 
\subsection{Monomial potential}
As  first case, we consider models characterized by the monomial potential 
\begin{equation}
V(\varphi)=\lambda\varphi^{n},
\label{60}
\end{equation}
where $\lambda$ is the parameter of the model with a mass scale. This class  of  effective potentials works in  particle physics and describes the interaction with other fields. The number $n$ is usually a positive integer \textit{e.g.} $n=2$ corresponds to the well-known potential $V=\frac{1}{2}M^{2}\varphi^{2}$ where $M$ is the mass of the inflaton and in the presence of the non-minimal coupling term (NMC) $\xi\varphi^{2}R$, we deal with an effective mass $m_{eff}=\sqrt{M^{2}+\xi R}$. The case $n=4$ is related to  $V=\lambda\varphi^{4}$ and describes the  chaotic inflation, where   $\lambda$ is a self-interacting constant \cite{Linde:1983gd,Linde:2007fr}. The cases  $n>6$ seems excluded by the observations. Furthermore, one can find some fractional powers in the literature \textit{e.g} $n = 2/3$ and $n = 4/3$ that could arise in axion monodromy inflation \cite{Flauger:2009ab,McAllister:2008hb,Kaloper:2008fb,Kaloper:2011jz,Kaloper:2014zba}. Here, we consider the cases $n=1,2,2/3,4/3$. 

Using the slow-roll parameters (\ref{57}) and (\ref{58}), the relation $3H\dot{\varphi}\simeq-V'$ and the definition of the number of e-folds (\ref{59}),
we can find the spectral parameters (\ref{33}) of different powers of $n$ considered according to  the following considerations.
\subsubsection{$n=2$}
In the case  $n=2$, spectral index and tensor-to-scalar ratio are  
\begin{equation}
n_{s}\simeq1+\frac{9\alpha}{4\gamma\lambda N^{2}}-\frac{3(16N^{2}\gamma^{2}\lambda^{2}-81\alpha^{2})}{N\mathcal{A}^{2}},\hspace{1cm}r\simeq\frac{8(16N^{2}\gamma^{2}\lambda^{2}-81\alpha^{2})}{N\mathcal{A}^{2}},
\label{62}    
\end{equation}
where $\mathcal{A}=-9\alpha+4\gamma\lambda N$.
\subsubsection{$n=\frac{4}{3}$}
In the case  $n=\frac{4}{3}$, spectral index and tensor-to-scalar ratio are 
\begin{equation}
n_{s}\simeq1-\frac{2304\sqrt{3N}\sqrt[4]{(\gamma\lambda)^{9}}}{\sqrt{\mathcal{A}}(-27\alpha+\sqrt{\mathcal{A}})^{2}}-12\sqrt{\frac{3}{N}}\sqrt[4]{(\gamma\lambda)^{3}}\frac{\Big(64\sqrt{(\gamma\lambda)^{3}}N\sqrt{\mathcal{A}}-243\alpha^{2}\sqrt{\mathcal{A}}+6561\alpha^{3}\Big)}{\mathcal{A}(-27\alpha+\sqrt{\mathcal{A}})^{2}},
\label{65}
\end{equation}
\subsubsection{$n=1$}
In the case  $n=1$, spectral index and tensor-to-scalar ratio are obtained as
\begin{equation}
n_{s}\simeq1-\frac{4\gamma^{2}\lambda^{2}}{\sqrt{\mathcal{A}}(-3\alpha+\sqrt{\mathcal{A}})^{2}}+\frac{4\gamma^{2}\lambda^{2}\Big(-2\gamma\lambda\sqrt[3]{16N^{2}\gamma\lambda}-27\alpha^{2}+9\alpha\sqrt{\mathcal{A}}\Big)}{3\mathcal{A}(-3\alpha+\sqrt{\mathcal{A}})^{3}},\hspace{1cm}
r\simeq\frac{32\gamma^{2}\lambda^{2}}{3\sqrt{\mathcal{A}}(-3\alpha+\sqrt{\mathcal{A}})^{2}},
\label{61}    
\end{equation}
where $\mathcal{A}=\gamma\lambda\sqrt[3]{16N^{2}\gamma\lambda}+9\alpha^{2}$.
\subsubsection{$n=\frac{2}{3}$}
For $n=\frac{2}{3}$, spectral index and tensor-to-scalar ratio are  
\begin{equation}
n_{s}\simeq1-\frac{357}{5}\sqrt[5]{\frac{(\gamma\lambda)^{9}}{N^{2}}}\frac{1}{\sqrt{\mathcal{A}}(-9\alpha+\sqrt{\mathcal{A}})^{2}}-\frac{12.25}{5}\sqrt[5]{\frac{(\gamma\lambda)^{3}}{N^{4}}}\frac{\Big(\sqrt{\mathcal{A}}\big(8\sqrt[5]{(\gamma\lambda)^{6}(90N)^{2}}+81\alpha^{2}\big)-54\alpha\sqrt[5]{(\gamma\lambda)^{6}(90N)^{2}}-729\alpha^{3}\Big)}{\mathcal{A}(-9\alpha+\sqrt{\mathcal{A}})^{2}},
\label{63}
\end{equation}
\begin{equation}
r\simeq\frac{952}{5}\sqrt[5]{\frac{(\gamma\lambda)^{9}}{N^{2}}}\frac{1}{\sqrt{\mathcal{A}}(-9\alpha+\sqrt{\mathcal{A}})^{2}},
\label{64}    
\end{equation}
where $\mathcal{A}=4\sqrt[5]{(\gamma\lambda)^{6}(90N)^{2}}+81\alpha^{2}$.

\begin{figure*}[!hbtp]
	\centering
	\includegraphics[width=.48\textwidth,keepaspectratio]{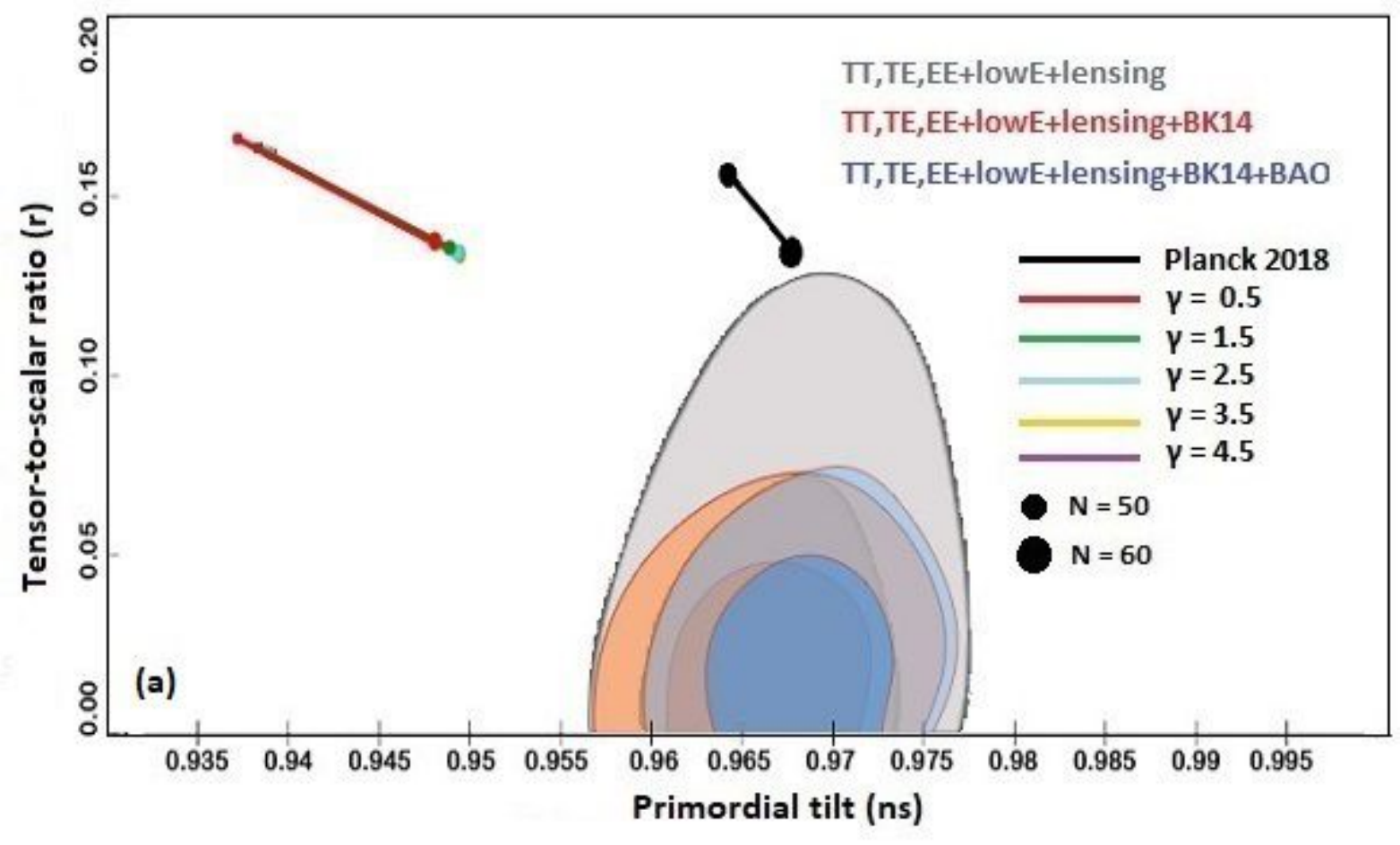}
	\includegraphics[width=.48\textwidth,keepaspectratio]{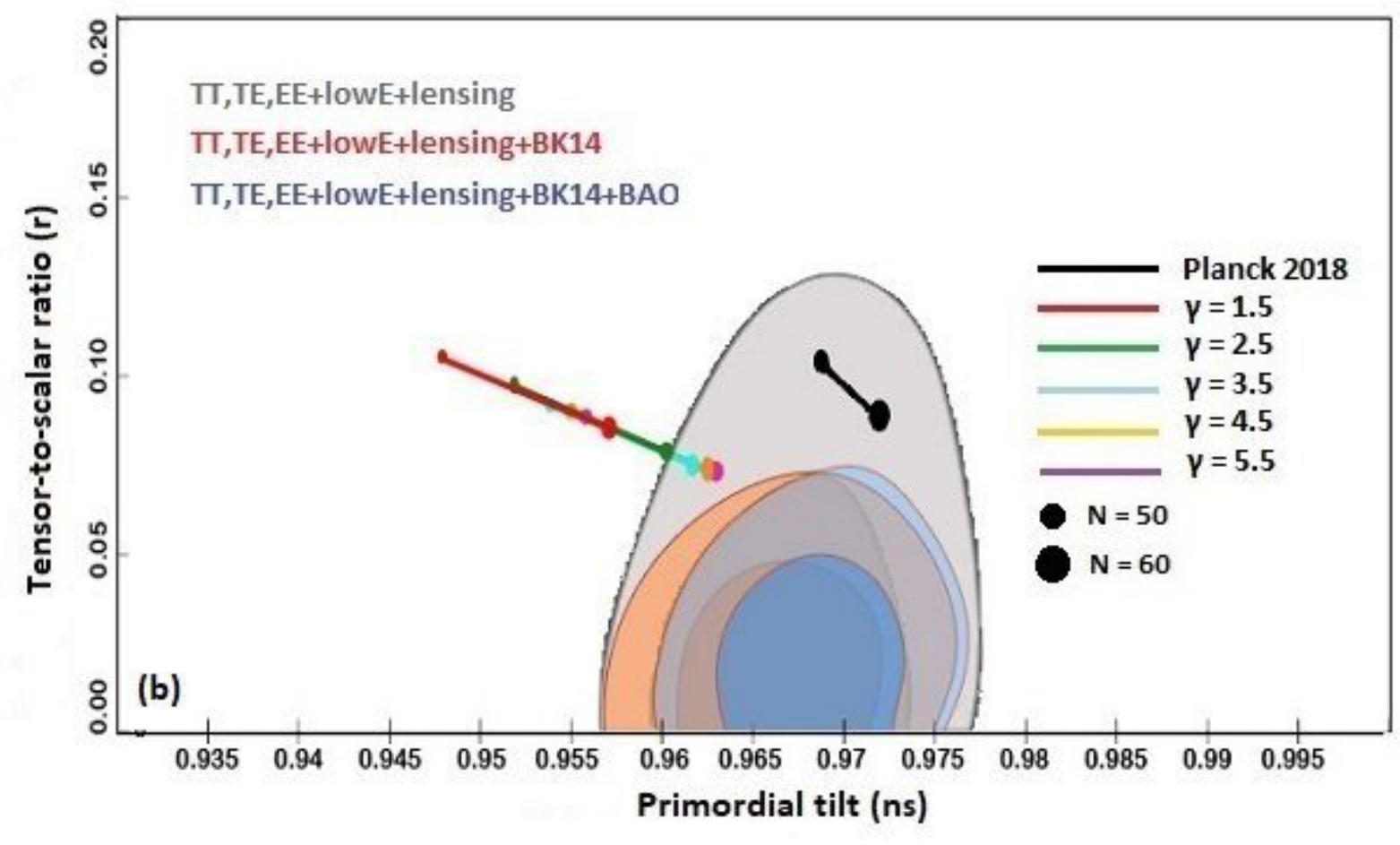}
	\includegraphics[width=.48\textwidth,keepaspectratio]{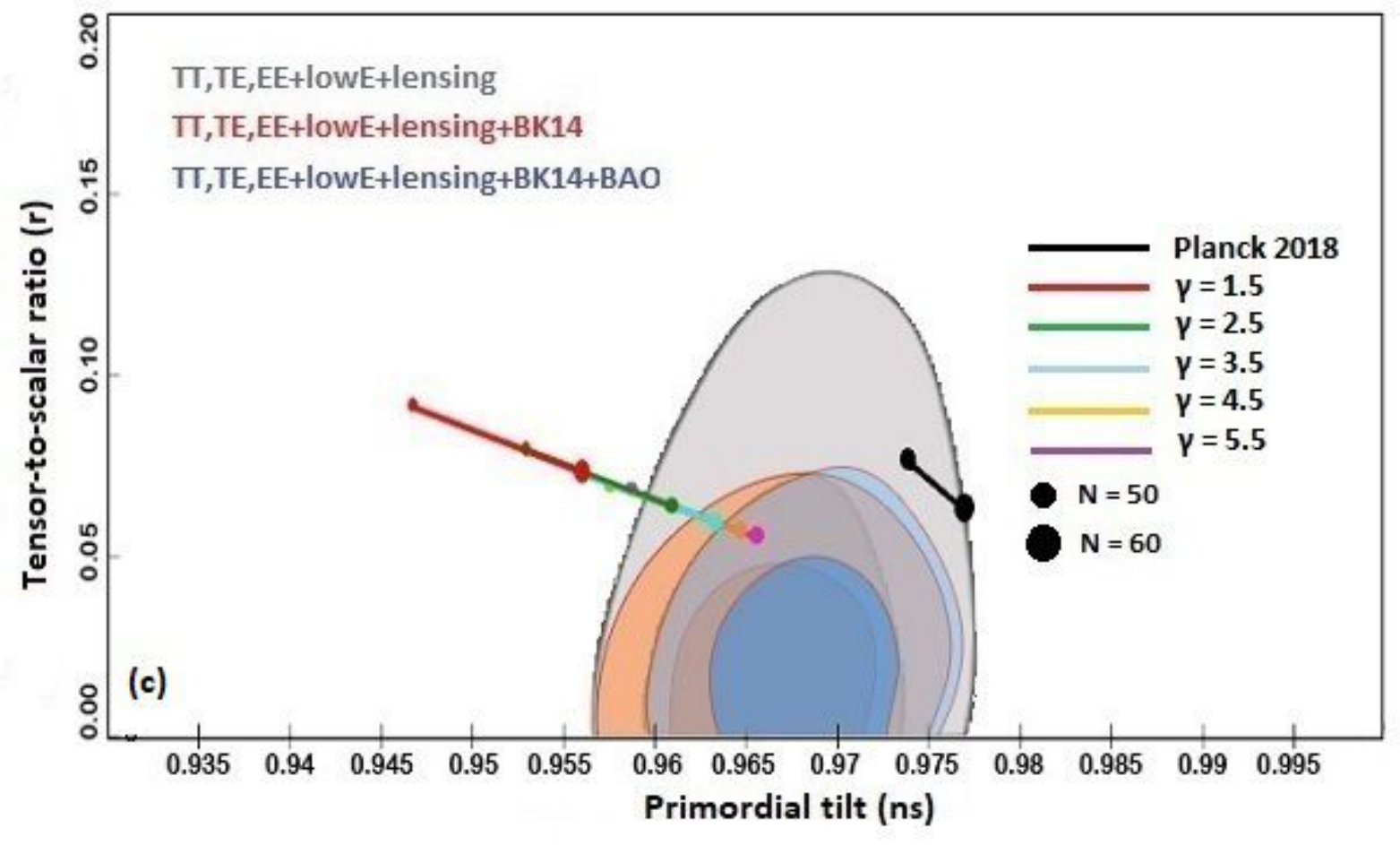}
    \includegraphics[width=.48\textwidth,keepaspectratio]{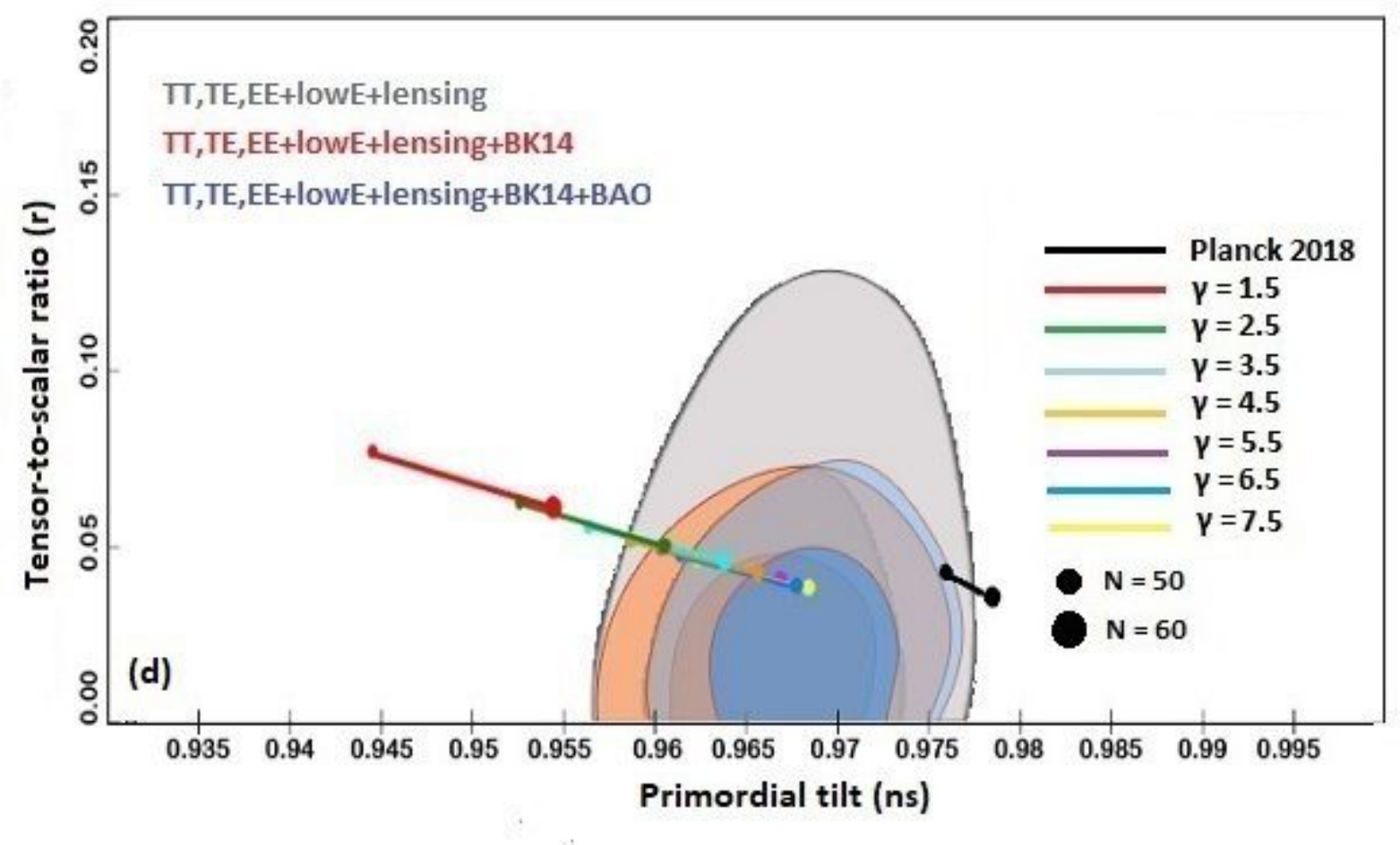}
	\caption{The marginalized joint 68\% and 95\% CL regions for $n_{s}$ and $r$ at $k = 0.002$ Mpc$^{-1}$ from  Planck alone and in combination with BK14 or BK14+BAO data \cite{cmb} and the $n_{s}-r$ constraints on the model (\ref{60}). \textit{(a)} Case of $n=2$ for $\gamma=0.5 M_{pl}^{-2}$ to $4.5 M_{pl}^{-2}$  when $\lambda=1.5 M_{pl}^{2}$. \textit{(b)} Case of $n=4/3$ for $\gamma=1.5 M_{pl}^{-8/3}$ to $5.5 M_{pl}^{-8/3}$  when $\lambda=1.5 M_{pl}^{8/3}$. \textit{(c)} Case of $n=1$ for $\gamma=1.5 M_{pl}^{-3}$ to $5.5 M_{pl}^{-3}$  when $\lambda=1.5 M_{pl}^{3}$. \textit{(d)} Case of $n=2/3$ for $\gamma=1.5 M_{pl}^{-10/3}$ to $7.5 M_{pl}^{-10/3}$  when $\lambda=1.5 M_{pl}^{10/3}$. All panels are plotted for $\alpha=1$.}
	\label{fig1}
\end{figure*}
\begin{equation}
r\simeq\frac{6144\sqrt{3N}\sqrt[4]{(\gamma\lambda)^{9}}}{\sqrt{\mathcal{A}}(-27\alpha+\sqrt{\mathcal{A}})^{2}},
\label{66}    
\end{equation}
where $\mathcal{A}=192\sqrt{(\gamma\lambda)^{3}}N+729\alpha^{2}$.

Now, let us compare the obtained results with the CMB anisotropies observations. In Fig.\ref{fig1}, we show the $n_{s}-r$ constraints coming from the marginalized
joint 68\% and 95\% CL regions of the Planck 2018 in combination with BK14+BAO data on the monomial model (\ref{60}) studied in the context of $f(Q)$ gravity. Panels show the prediction of the model in the cases of $n=2,4/3,1,2/3$ compared with their counterparts in Planck 2018 release (black line) for $N=50$ (small circle) and $N=60$ (big circle). In panel(a), we present the behavior of the model in the case of $n=2$ for $\gamma=0.5M_{pl}^{-2}$ to $4.5M_{pl}^{-2}$ when $\lambda=1.5M_{pl}^{2}$ and $\alpha=1$.  Generally, the monomial potential with $n=2$ is completely ruled out by the Planck observations since its values for $n_{s}$ and $r$ are situated out of the observational regions (black line). As we can see, our findings about $n_{s}$ and $r$ correspond to different values of $\gamma$. This fact  tells us that this result is still valid for the case of $n=2$ in $f(Q)$ gravity (colorful lines). In the case  $n=4/3$, plotted for $\gamma=1.5M_{pl}^{-8/3}$ to $5.5M_{pl}^{-8/3}$ when $\lambda=1.5M_{pl}^{8/3}$ and $\alpha=1$ in panel (b), Planck predicts the values of $n_{s}\sim0.97$ and $r\sim0.1$ situated in the Planck  region only at the 68\% CL while prediction of our model in $f(Q)$ is not fully consistent with Planck data. The panel shows that, for $\gamma\leq2.5$, the obtained values of $n_{s}$ and $r$ are not in good agreement with the observations. However, for $\gamma>2.5$, in the case of $N=60$, our results are compatible with Planck prediction (black line) even with more observational values of the spectral parameters. Note that by considering $\gamma>5.5$, we get more favored results since the values of $n_{s}$ and $r$ are placed in the regions of Planck data in  combination with BK14 and BAO data at  68\% CL. Panel (c) corresponds to the case  $n=1$ for $\gamma=1.5M_{pl}^{-3}$ to $5.5M_{pl}^{-3}$ when $\lambda=1.5M_{pl}^{3}$ and $\alpha=1$. From Planck 2018 (black line), we find $n_{s}\sim0.975$ and $r\sim0.07$ that are situated in the Planck alone at the 68\% CL. The panel reveals that the cases $\gamma<1.5$ are completely excluded while the cases $\gamma\geq 2.5$ present  acceptable observational  values of $n_{s}$ and $r$, in particular, for $N=60$.
\begin{figure*}[!hbtp]
	\centering
	\includegraphics[width=.55\textwidth,keepaspectratio]{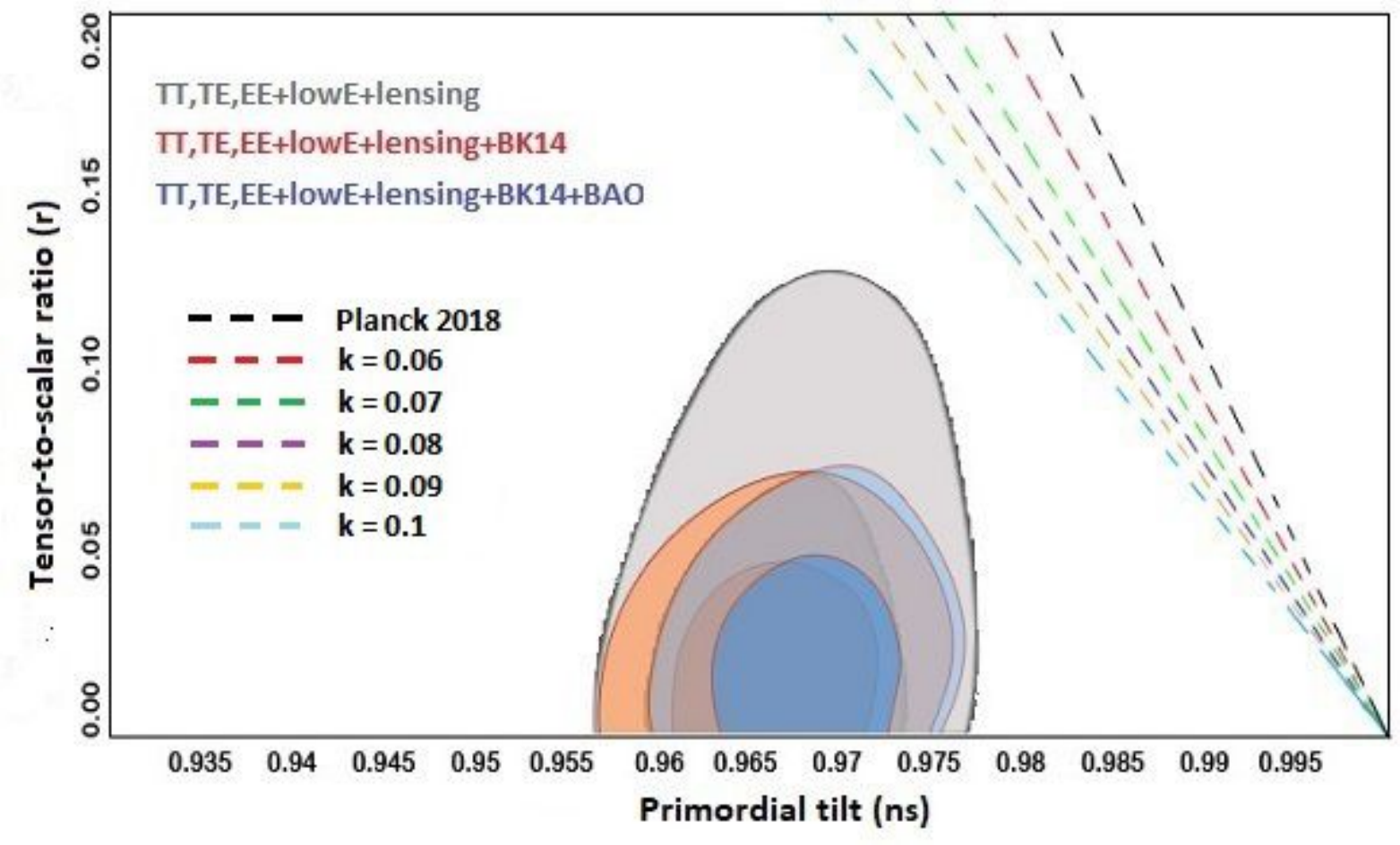}
	\caption{The marginalized joint 68\% and 95\% CL regions for $n_{s}$ and $r$ at $\kappa = 0.002$ Mpc$^{-1}$ from Planck alone and in combination with BK14 or BK14+BAO data \cite{cmb} and the $n_{s}-r$ constraints on the exponential model (\ref{67}) for $k=0.06M_{pl}^{-1}$ to $0.1M_{pl}^{-1}$ when $N=60$ and $\alpha\sim M_{pl}^{2}$.}
	\label{fig2}
\end{figure*}
Moreover, we find that, for $\gamma>5.5$, our model predicts  values of $n_{s}$ and $r$ more consistent with observations due to their consistency with the combined Planck with BK14 and BAO datasets at  68\% CL. In panel (d), we present the comparison of the monomial potential in the case of $n=2/3$ with Planck data for $\gamma=1.5M_{pl}^{-10/3}$ to $7.5M_{pl}^{-10/3}$ when $\lambda=1.5M_{pl}^{10/3}$ and $\alpha=1$. For $N=50$, the Planck 2018 (black line) predicts a favored value of $n_{s}$ and also $r$ consistent with the Planck data in a combination of BK14 and BAO data at  68\% CL while it is ruled out for $N=60$. Also, the panel presents that our model is in good agreement with the observations for $\gamma>2.5$ in both cases $N=50$ and $N=60$ since they are placed in the region of Planck in combination with BK14 and BAO datasets in both 65\% and 95\% CL. Analogous to the previous cases, for higher values of $\gamma$, the results of our model are more consistent with the observational data.

\subsection{Power-law inflation}
Another interesting case is the exponential potential
\begin{equation}
V=V_{0}e^{-k\varphi},
\label{67} 
\end{equation}
associated to the power-law inflation where the  scale factor of the universe evolves as $a(t)\propto t^{q}$, $q>1$ \cite{Abbott:1984fp,Lucchin:1984yf,Sahni:1990tx}. This class of inflationary models is usually ruled out by the Planck data \cite{cmb}. Here $k$ has the dimension $[mass]^{-1}$ and $V_{0}$ refers to the energy scale with the dimension of $[mass]^{4}$. 

Combining the slow-roll parameters (\ref{57}) and (\ref{58}), the reduced Klein-Gordon equation $3H\dot{\varphi}\simeq-V'$ and the number of e-folds (\ref{59}), the spectral parameters (\ref{33}) of the model are  $\gamma$-independent and driven as
\begin{eqnarray}
&\!&\!n_{s}\simeq-\frac{2}{N\mathcal{A}^{3}(\alpha Nk^{2}-\mathcal{A})^{2}}\bigg\{N\mathcal{A}\Big(N^{4}\alpha^{4}k^{8}+2(3-\alpha k^{2})\alpha^{3} k^{6}N^{3}+2(7-4\alpha k^{2})\alpha^{2} k^{4}N^{2}+\nonumber\\&\!&\!
+4(4-3\alpha k^{2})\alpha k^{2}N-6\alpha k^{2}+8\Big)+\mathcal{A}^{2}\Big(2+N^{4}k^{6}\alpha^{3}+2(2-\alpha k^{2})\alpha^{2} k^{4}N^{3}+4N^{2}k^{2}\alpha+4\alpha Nk^{2}\Big)\bigg\},
\label{68}    
\end{eqnarray}
\begin{equation}
r\simeq\frac{32(\mathcal{A}-1)^{2}}{N^{2}k^{2}\mathcal{A}\Big(-\alpha Nk^{2}+\mathcal{A}\Big)^{2}},
\label{69}    
\end{equation}
where $\mathcal{A}=\alpha Nk^{2}+2$.

In Fig.\ref{fig2}, we present the $n_{s}-r$ constraints coming from the marginalized joint 68\% and 95\% CL regions of the Planck 2018 in combination with BK14+BAO data on the exponential model (\ref{67}) in the context of $f(Q)$ gravity. The figure shows the predictions of the model in case of $k=0.06M_{pl}^{-1}$ to $0.1 M_{pl}^{-1}$ compared with its counterpart in Planck 2018 release (black line) when $N=60$ and $\alpha\sim M^{2}$. From the figure, we realize that the obtained values of $n_{s}$ and $r$ associated with different values of $k$ are situated in observationally disfavoured regions. Hence, the exponential potential in the $f(Q)$ theory is not in good agreement with the CMB observations. This result is consistent with the Planck 2018 prediction that rules out power-law inflation (black dashed line). 
\begin{figure*}[!hbtp]
	\centering
	\includegraphics[width=.48\textwidth,keepaspectratio]{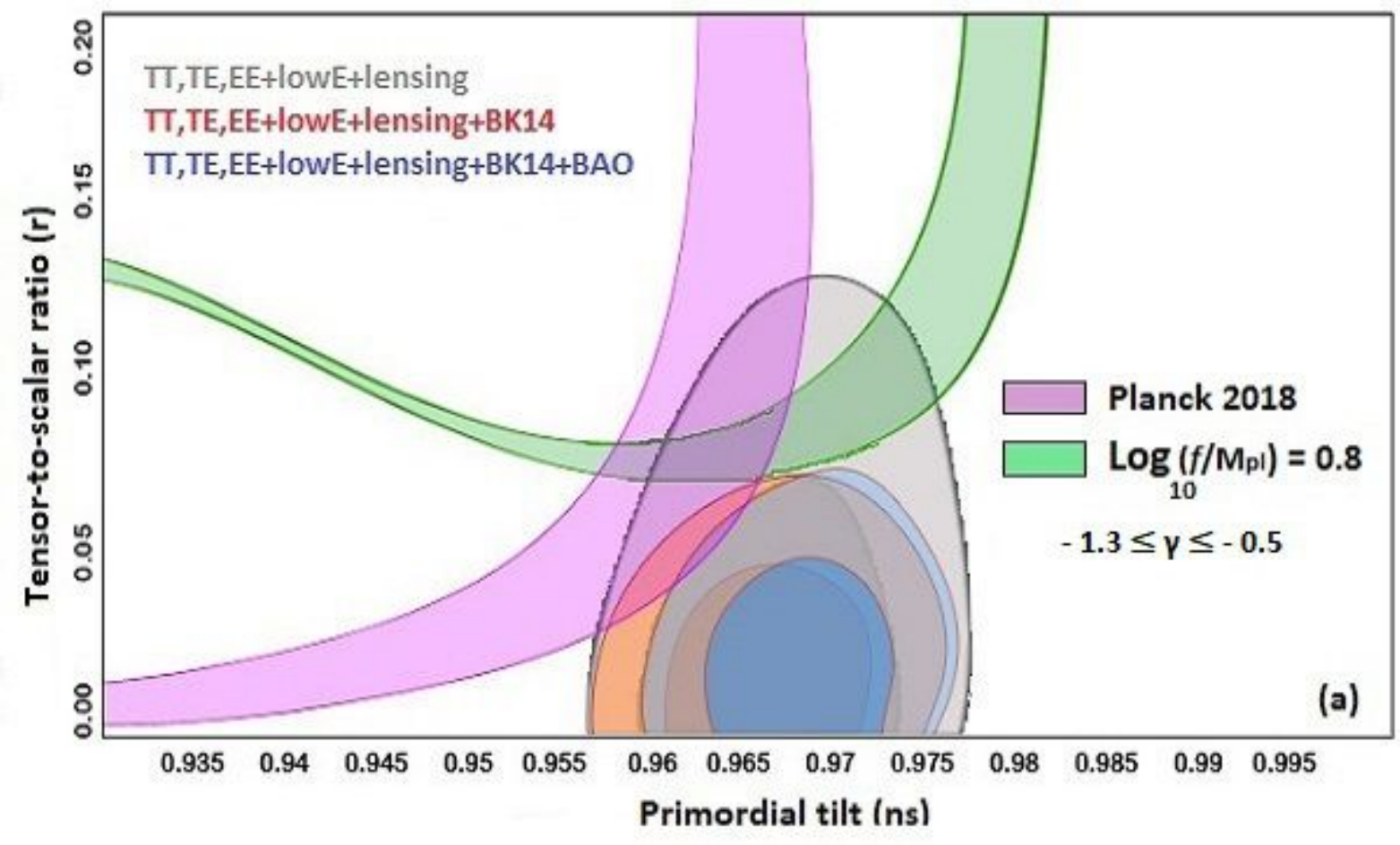}
	\includegraphics[width=.48\textwidth,keepaspectratio]{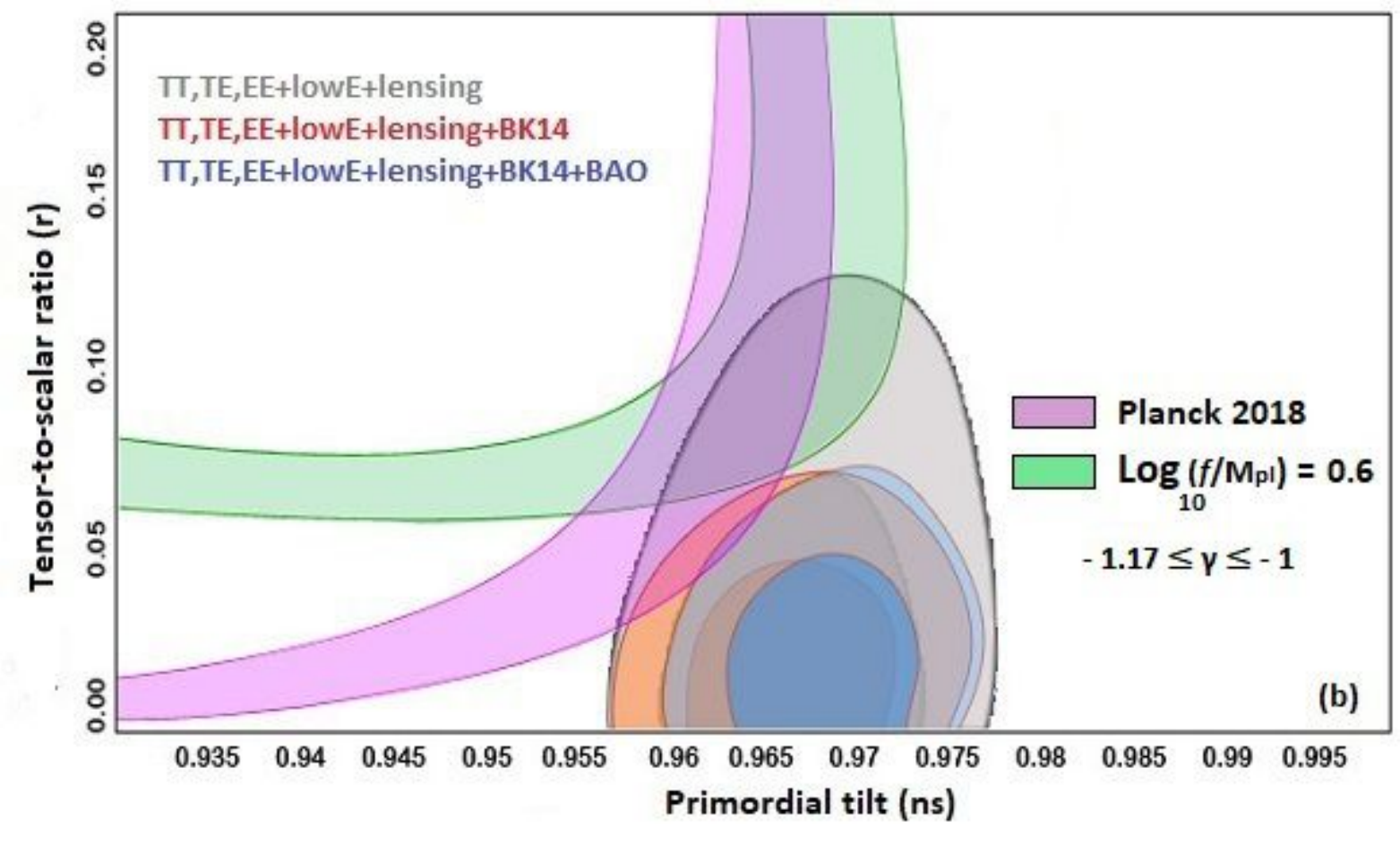}
	\includegraphics[width=.48\textwidth,keepaspectratio]{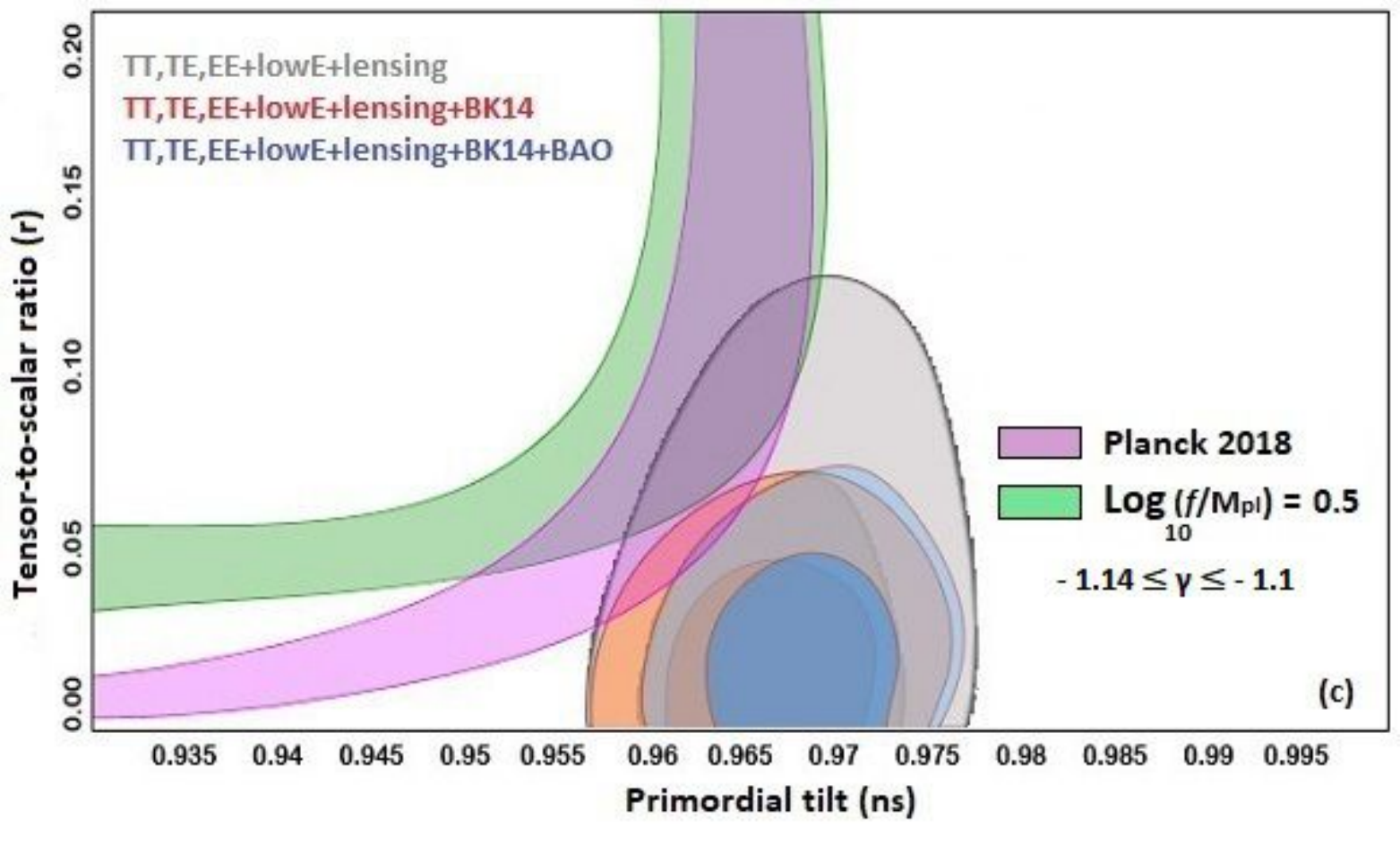}
	\includegraphics[width=.48\textwidth,keepaspectratio]{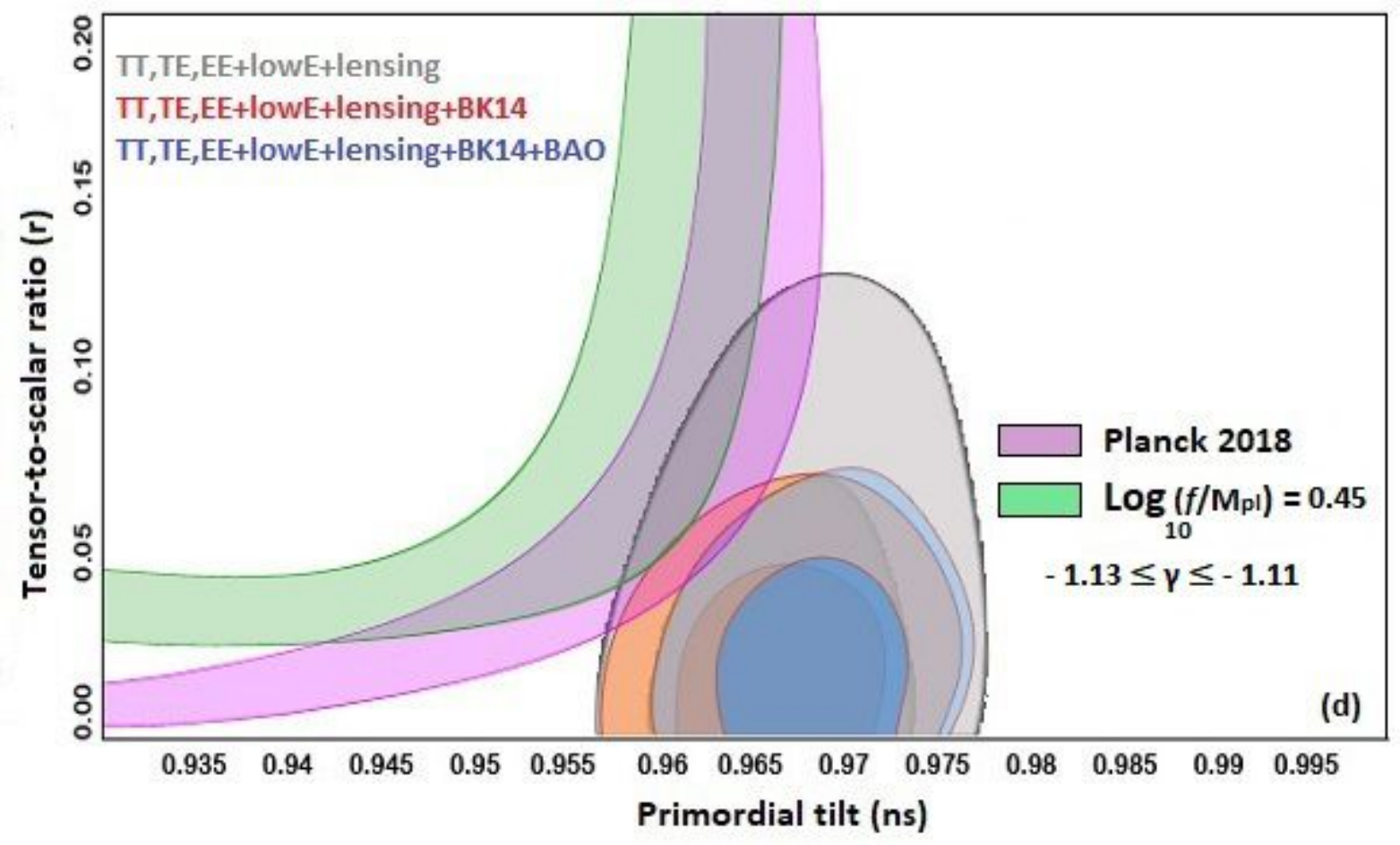}
	\caption{The marginalized joint 68\% and 95\% CL regions for $n_{s}$ and $r$ at $k = 0.002$ Mpc$^{-1}$ from  Planck alone and in combination with BK14 or BK14+BAO data \cite{cmb} and the $n_{s}-r$ constraints on the natural model (\ref{70}). \textit{(a)} Case of $\log_{10}(f/M_{pl})=0.8$ for allowed value of $-1.3\leq\gamma\leq-0.5$. \textit{(b)} Case of $\log_{10}(f/M_{pl})=0.6$ for allowed value of $-1.17\leq\gamma\leq-1$. \textit{(c)} Case of $\log_{10}(f/M_{pl})=0.5$ for allowed value of $-1.14\leq\gamma\leq-1.1$. \textit{(d)} Case of $\log_{10}(f/M_{pl})=0.45$ for allowed value of $-1.13\leq\gamma\leq-1.11$. All panels are plotted for $\alpha\sim M_{pl}^{2}$.}.
	\label{fig3}
	\end{figure*}
\subsection{Natural inflation}
A reliable potential in the standard inflationary paradigm requires proposing a mechanism to remove the flatness problem as one of the main shortcomings of the hot big bang theory. Natural inflation (NI) overcomes this problem using a Pseudo-Nambu-Goldstone boson as inflaton with a flat potential \cite{Adams:1992bn,Freese:1990rb}
\begin{equation}
V(\varphi)=V_{0}\Big(1+\cos(\frac{\varphi}{f})\Big),
\label{70}
\end{equation}
in which a global $U(1)$ symmetry is spontaneously
broken at scale $f$, with  explicit soft symmetry breaking
at a lower scale $\Lambda$. Here, $V_{0}=\Lambda^{4}$ with $\Lambda\sim m_{GUT}$ and $f\sim m_{pl}$. For the small angle approximation $f\ll\varphi$, the potential reduces to $V\simeq2\Lambda^{4}$ while for large angle approximation $f\gg
m_{pl}$, its recovers chaotic inflation ${\displaystyle V(\psi)=\frac{m^{2}\psi^{2}}{2}}$ where $\psi=\varphi-\sigma$ and
$\sigma=constant$.

Using the slow-roll parameters (\ref{57}) and (\ref{58}) of the model, the expression $3H\dot{\varphi}\simeq-V'$ and the number of e-folds (\ref{59}), one can obtain the spectral parameters (\ref{33}) of the model as
\begin{eqnarray}
&\!&\!n_{s}=1-\frac{12\gamma^{2}V_{0}^{2}(1-\mathcal{A}^{2})}{3f^{2}\sqrt{4V_{0}(1+\mathcal{A})\gamma+9\alpha^{2}}\Big(-3\alpha+\sqrt{4V_{0}(1+\mathcal{A})\gamma+9\alpha^{2}}\Big)^{2}}-\nonumber\\&\!&\!-\frac{2\gamma V_{0}}{3f^{2}\Big(4V_{0}(1+\mathcal{A})\gamma+9\alpha^{2}\Big)\Big(-3\alpha+\sqrt{4V_{0}(1+\mathcal{A})\gamma+9\alpha^{2}}\Big)^{2}}\Bigg\{\gamma V_{0}+\Big(9\alpha-2\sqrt{4V_{0}(1+\mathcal{A})\gamma+9\alpha^{2}}\Big)+\nonumber\\&\!&\!+\Big(4V_{0}(1+\mathcal{A})\gamma+9\alpha^{2}\Big)\Big(-3\alpha+\sqrt{4V_{0}(1+\mathcal{A}))\gamma+9\alpha^{2}}\Big)\Bigg\},
\label{71}
\end{eqnarray}
\begin{equation}
r=\frac{32\gamma^{2}V_{0}^{2}(1-\mathcal{A}^{2})}{3f^{2}\sqrt{4V_{0}(1+\mathcal{A})\gamma+9\alpha^{2}}\Big(-3\alpha+\sqrt{4V_{0}(1+\mathcal{A})\gamma+9\alpha^{2}}\Big)^{2}},
\label{72}    
\end{equation}
where $\mathcal{A}=1-e^{\frac{4\gamma V_{0}N}{9\alpha f^{2}}}$.

Fig.\ref{fig3} shows the $n_{s}-r$ constraints coming from the marginalized
joint 68\% and 95\% CL regions of the Planck 2018 in combination with BK14+BAO data on NI (\ref{70}) in the regime of $f(Q)$ theory. In the figure, we find the predictions of the model for the allowed values of $\gamma$ in four cases of $\log_{10}(f/M_{pl})$ in comparison with the result released in Planck 2018 for NI (pink color). The panel (a) is drawn for the case of $\log_{10}(f/M_{pl})=0.8$ for the allowed values of $-1.3\leq\gamma\leq-0.5$ in which there is a minimum overlap between NI in $f(Q)$ gravity (green color) and NI in the Planck 2018 release (pink color). By decreasing the value of $\log_{10}(f/M_{pl})$ to $0.6$ in panel (b), one can realize that more regions of two models (\textit{i.e.} NI in Planck and $f(Q)$ gravity) are overlapped together, in particular, for $r>0.05$. In this case, the allowed range of $\gamma$ is reduced to $-1.17\leq\gamma\leq-1$. In panel (c), we decline the value of $\log_{10}(f/M_{pl})$ to $0.5$ for the allowed values of $-1.14\leq\gamma\leq-1.1$. As we can see, the overlapping starts from $r>0.04$ and keeps going for bigger values of $r$. For a more interesting case, we consider $\log_{10}(f/M_{pl})=0.45$ in a narrow range of  $-1.13\leq\gamma\leq-1.11$ in which two NI models behave almost the same with a maximum overlap beginning from $r=0.03$. Note that all panel of Fig.\ref{fig3} are plotted for $N=50-60$ and $\alpha\sim M^{2}$.

\section{Discussion and Conclusions}\label{con}

In this paper, we discussed the possibility to realize cosmological slow-roll inflation in the framework of extended non-metric theories of gravity where a generic functions $f(Q)$ of the non-metricity scalar $Q$ have been  considered. 

The analysis has been developed after discussing the conformal transformations in non-metric gravity where further features emerge with respect to the analogue formulation in metric theories of gravity  like $f(R)$. 

In the latter case, it is straightforward to pass from the Jordan frame to the Einstein frame: the further degrees of freedom related to any extended theory can be easily recast as a Hilbert-Einstein Lagrangian plus  a scalar-field Lagrangian where kinetic  and potential terms are well distinguished.

In the $f(Q)$ case, kinetic and potential terms can be clearly divided only in the strong field regime while they mix in the weak field limit. This means that when inflation is going to end up, not only the standard kinetic term starts to work but also the mixed coupling comes into the game. In other words, as soon as the system transits into the reheating phase, also conformal invariance is going to break as we will discuss below.

In this perspective, we analysed the potential slow-rolling (PSR) and the Hubble slow-rolling (HSR). In the first case, we assumed that conformal transformations are working and then the Einstein frame is restored. As a consequence, the $f(Q)$ function gives rise to the inflationary potential and it is possible to compare the various forms of potential with the Planck 2018, BK14 and BAO datasets. In particular, we analysed $f(Q)$ gravity with quadratic corrections (and compared it with $R^2$ Starobinsky model), a generic power-law correction in $Q$, and, finally,  a logarithmic correction to the $Q^2$ model. In  all these cases, it is possible to constrain inflationary parameters and select the marginalized joint 68\% and 95\% CL regions in the space parameters. Results in PSR approach can be summarized as follows:
\begin{itemize}

\item 
Concerning the $Q^{\theta}$ model, we found that the case of $\theta=2$ is compatible with the CMB observations coming from Planck+BK14+BAO with a larger value of tensor-to-scalar ratio $r$ with respect to the $R^{2}$ Starobinsky model. On the other hand, by going beyond the case of $\theta=2$, we realized that the observations allow us to consider just a tiny deviation from $\theta=2$ of the order of $10^{-2}$. This result is in agreement with the $f(Q)$ model proposed in Ref.\cite{BeltranJimenez:2019tme} for inflation. For the Planck alone dataset  and also its combination with the BK14, we obtained the observationally allowed range of $2.02\leq\theta\leq2.05$ and $2.03\leq\theta<2.05$ at the $68\%$ and $95\%$ C.L., respectively. 
For the combined Planck+BK14+BAO observations of the CMB anisotropy, the observational constraints reduce to $2.02\leq\theta\leq2.05$ and $2.03\leq\theta\leq2.04$ at the $68\%$ and $95\%$ C.L., respectively. Note that the value of the tensor-to-scalar ratio $r$ for the case of $\theta\neq 2$ is smaller than the case of $\theta=2$ and also the $R^{2}$ Starobinsky model. 

\item 
As a generalization of the $Q^{2}$ model, we considered the logarithmic corrected model  $f(Q)=Q+\xi Q^{2}+\upsilon Q^{2}\ln Q$. From the Planck alone datasets and its combination with the BK14, we found the observational constraint $-0.03\xi<\upsilon<-0.015\xi$ at the 68\% C.L. and $-0.025\xi\leq\upsilon\leq-0.02\xi$ at the 95\% C.L. For the Planck+BK14+BAO datasets, the constraints are reduced to $-0.025\xi\leq\upsilon<-0.015\xi$ and $-0.025\xi\leq\upsilon\leq-0.02\xi$ at the 68\% C.L. and 95\% C.L., respectively.

\end{itemize}

A similar analysis have been developed for the HSR approach. In this case, we remain in the Jordan frame and study monomial, power-law and natural inflation scenarios. Here the $f(Q)$ background is fixed as $f(Q)=\alpha Q+\beta Q^{m}$ where $\alpha$ and $\beta$ are the parameters of the model. Moreover, $m$ is a dimensionless parameter and could be considered as $m<1$ for the low-curvature DE regime  and as $m>1$ for the inflationary high curvature regime \cite{BeltranJimenez:2019tme}. Then, the inflationary analysis can be performed accordingly. The cosmological equations results modified with respect to the GR (or STEGR) ones and the slow-roll parameters $\epsilon$ and $\eta$ are given directly by the evolution of the Hubble parameter $H$ and its derivatives.  Also in  this case, it is possible to constrain inflationary parameters and select marginalized joint 68\% and 95\% CL regions using Planck 2018, BK14 and BAO  datasets excluding or retaining parameter regions. Our results in HSR approach can be summarized as follows:
\begin{itemize}

\item For the monomial potential $V\propto\varphi^{n}$ in the context of $f(Q)$ gravity, we found that the results are almost in agreement with the results coming from the Planck 2018 for all powers of $n$. In the case of $n=2$, the obtained values of $n_{s}$ and $r$ are not compatible with their observational values. In the case of $n=\frac{4}{3}$, for $\gamma>2.5$ and $N=60$, our results are compatible with the observations with more acceptable values of the tensor-to-scalar ratio with respect to the Planck 2018. In the case of $n=1$, we found that the $\gamma<1.5$ are fully ruled out by the observations while the range $\gamma\geq2.5$ is in good agreement with the observations for $N=60$. In the case of $n=2/3$, our results are compatible with the observations for $\gamma>2.5$ in both cases $N=50$ and $N=60$. 
\item 
For the power-law inflation with the potential $V\propto e^{-k\varphi}$ in $f(Q)$ theory, the obtained values of $n_{s}$ and $r$ for different values of $k$ are excluded by the observations in analogy with the results coming from  Planck 2018.
\item For the natural inflation (NI) with the potential $V(\varphi)\propto1+\cos(\frac{\varphi}{f})$ introduced in $f(Q)$ regime, we found the observational constraint $-1.3\leq\gamma\leq-0.5$ for the case of $\log_{10}(f/M_{pl})=0.8$. By choosing the case $\log_{10}(f/M_{pl})=0.6$, we obtained the observational constraint $-1.17\leq\gamma\leq-1$ which is more compatible with the NI in the Planck 2018. The constraint could be reduced to $-1.14\leq\gamma\leq-1.1$ by considering $\log_{10}(f/M_{pl})=0.5$. As the most compatible case with the NI of Planck 2018 data, we found the constraint $-1.13\leq\gamma\leq-1.11$ for $\log_{10}(f/M_{pl})=0.45$. 
\end{itemize}

From a methodological  point of view, the HSR approach could be more reliable because it is  directly  based on the evolution of observables like $H$ and its derivatives. This fact allows an immediate  comparison with data and the fact we are not performing  conformal transformations avoids  any interpretative issue related with variables.
 
In the case of PSR, the potential has to be "recovered" from a  conformal transformation. This fact allows to compare models with analogue ones in the minimal coupling regime where Einstein gravity is improved with scalar field dynamics. As we have shown, in the case of $f(Q)$,  the validity of conformal transformations strictly depends on the field regime. As discussed above, the presence of the third term in the action \eqref{26} gives rise to an additional scalar-non-metricity coupling   which cannot be removed by a standard conformal transformation. This means that $f(Q)$ models  are not dynamically equivalent to the STEGR action plus a scalar field via a conformal transformation like in the metric case of $f(R)$ gravity where GR plus a scalar field is fully recovered.  In the $f(Q)$ case, the standard  conformal structure is recovered in high-energy regime while it breaks in the weak field limit. This fact is not immediately  comparable  with observations because "observables" cannot be easily selected  but it can be relevant from a conceptual point of view. In fact,  the emergence of the third term in \eqref{26} could be the signal of the exit from inflation pointing out a symmetry breaking leading to the post-inflationary universe. From a general  viewpoint,  the fact that isometries are not requested at the foundation of  non-metric gravity \cite{Capozziello:2022zzh} could an advantage related to the intrinsic meaning of inflationary mechanism:  here "scales" are changing and  they do not necessarily have to change  "isometrically". Furthermore, also the Equivalence Principle is not required "a priori" so also possible violations in some phase transition could be considered in $f(Q)$ dynamics.  
 
 In a forthcoming paper, these aspects will be deeply investigated and confronted with observations.

\section{Acknowledgments}
SC acknowledges the support of Istituto Nazionale di Fisica Nucleare (INFN) {\it iniziative specifiche} QGSKY and Moonlight2.

\bibliographystyle{ieeetr}
\bibliography{biblo}
\end{document}